\documentclass[journal,twocolumn]{IEEEtran}
\IEEEoverridecommandlockouts

\usepackage{bbding}
\usepackage{threeparttable}
\usepackage{hyperref}
\usepackage{rotating}
\usepackage{multirow}
\usepackage{cite}

\usepackage{subfigure}
\usepackage{amsfonts}
\usepackage{amsmath,amssymb}
\usepackage[ruled,vlined,linesnumbered]{algorithm2e}

\usepackage{algorithmicx}
\usepackage{graphicx}
\usepackage{textcomp}
\usepackage{xcolor}
\usepackage{booktabs}
\usepackage{color}
\def\BibTeX{{\rm B\kern-.05em{\sc i\kern-.025em b}\kern-.08em
		T\kern-.1667em\lower.7ex\hbox{E}\kern-.125emX}}
\begin{document}

\title{Learnware for CSI Feedback: \\Scene-specific Small Models Can Do Big}

\author{
\normalsize {Xiangyi~Li,
\IEEEmembership{\normalsize {Graduate Student Member,~IEEE}},
Jiajia~Guo, \IEEEmembership{\normalsize {Member,~IEEE}},
Chao-Kai~Wen, \IEEEmembership{\normalsize {Fellow,~IEEE}},\\
Xin Geng, \IEEEmembership{\normalsize {Senior Member,~IEEE}},
Shi~Jin, \IEEEmembership{\normalsize {Fellow,~IEEE}},
and Zhi-Hua~Zhou, \IEEEmembership{\normalsize {Fellow,~IEEE}}
}
\thanks{X.~Li, J.~Guo, and S.~Jin are with the National Mobile
Communications Research Laboratory, Southeast University, Nanjing, 210096, China (email: Xiangyi\_li@seu.edu.cn, jiajiaguo@seu.edu.cn, jinshi@seu.edu.cn). J. Guo is now with the Department of Electronic and Computer Engineering, The Hong Kong University of Science and Technology.}
\thanks{C.-K.~Wen is with the Institute of Communications Engineering, National Sun Yat-sen University, Kaohsiung 80424, Taiwan (e-mail: chaokai.wen@mail.nsysu.edu.tw).}
\thanks{X.~Geng is with the School of Computer Science and Engineering and the Key Laboratory of New Generation Artificial Intelligence Technology and Its Interdisciplinary Applications (Southeast University), Ministry of Education, Southeast University, Nanjing 211189, China (e-mail: xgeng@seu.edu.cn).}
\thanks{Z.-H.~Zhou is with the National
Key Laboratory for Novel Software Technology, and also with the School of
Artificial Intelligence, Nanjing University, Nanjing 210023, China (e-mail: zhouzh@lamda.nju.edu.cn).}
}

\maketitle

\begin{abstract}
Intelligent channel state information (CSI) feedback is essential for realizing the high capacity and spectral efficiency goals of future 6G systems, yet existing deep learning solutions face a trade‐off between model generalization and scenario‐specific performance. Large neural networks generalize well but incur high computational and tuning costs, while small models excel in particular environments but require repetitive costly end‐to‐end training for each base station (BS). To address these challenges, we introduce a model repository-based deployment framework in which a centralized AI data center maintains a catalog of scene‐specific CSI models. The repository is enhanced with a Learnware-based framework, where each model is associated with a \textit{specification} including semantic part (network architecture parameters) and statistical part (codebook‐fingerprint embeddings of training‐data distributions). A BS submits only its local statistical specifications to retrieve the most relevant pre‐trained model, \textcolor{black}{enhancing data privacy by avoiding raw CSI transmission} and drastically reducing retrieval latency and communication overhead. We further develop a data‐driven search strategy that matches codebook fingerprints to model performance, achieving over 90\% selection accuracy. In simulations, our scheme yields \textcolor{black}{18.8\% and 57.7\% performance improvements over the General Model in LOS and NLOS scenarios, respectively} while reducing local fine‐tuning \textcolor{black}{by up to 1000 samples} and 100 epochs. This Learnware‐based approach minimizes redundant training, maximizes model reuse, and supports rapid, \textcolor{black}{privacy‐enhancing} deployment of CSI feedback models.
\end{abstract}


\IEEEpeerreviewmaketitle

\section{Introduction}
\label{introduction}
\IEEEPARstart{A}{ccording} to the latest 6G vision released by the International Telecommunication Union (ITU) \cite{ITU-WP5D1}, future wireless communication systems will demand higher throughput, lower latency, greater reliability, massive connectivity, and more efficient spectrum utilization \cite{dang2020should, alhammadi2024artificial}. Massive multiple-input multiple-output (MIMO) arrays are essential to achieving these goals, offering spatial multiplexing gains that improve both capacity and spectral efficiency \cite{
marzetta2015massive}. In frequency division duplexing (FDD) MIMO systems, user equipment (UE) estimates the downlink channel state information (CSI) and feeds it back to the base station (BS) for precoding. Since transmitting full CSI entails substantial overhead, efficient compression and reconstruction mechanisms are critical.

Conventional CSI compression techniques, such as compressed sensing \cite{eltayeb2014compressive} and codebook-based methods \cite{Ra2015codebook}, provide theoretical guarantees but often underperform in complex, large-scale deployments \cite{guo2022overview}. Deep learning (DL)-based methods have emerged as effective alternatives by modeling CSI matrices as high-dimensional images and training autoencoders for end-to-end compression and reconstruction. This approach was first demonstrated in CsiNet \cite{wen2018deep}, which achieved significant reductions in feedback overhead while maintaining reconstruction accuracy. Subsequent works have incorporated convolutional structures, attention mechanisms, and lightweight network designs to further improve performance and reduce inference latency \cite{an2025deep, jiang2024digital, Chen2025CSI,liang2024toward, Shin2024Vector, Lin2024Scalable, Peng2024Learning-Based}. In parallel with these research advances, the 3rd Generation Partnership Project (3GPP) initiated AI-driven CSI feedback as a study item in Release 18 (R18) and subsequently advanced it to a work item in Release 20 \cite{3GPP38843, Lin2025Mobile}. More broadly, AI has been recognized as a key enabler of 6G and is being incorporated into air interface design across multiple 3GPP working groups \cite{Hoydis2021Toward, ye2024artificial, huang2025aiml}.

\begin{figure}
    \centering
    \setlength{\belowcaptionskip}{-1cm}
    \setlength{\abovecaptionskip}{0cm}
    \includegraphics[width=0.9\linewidth]{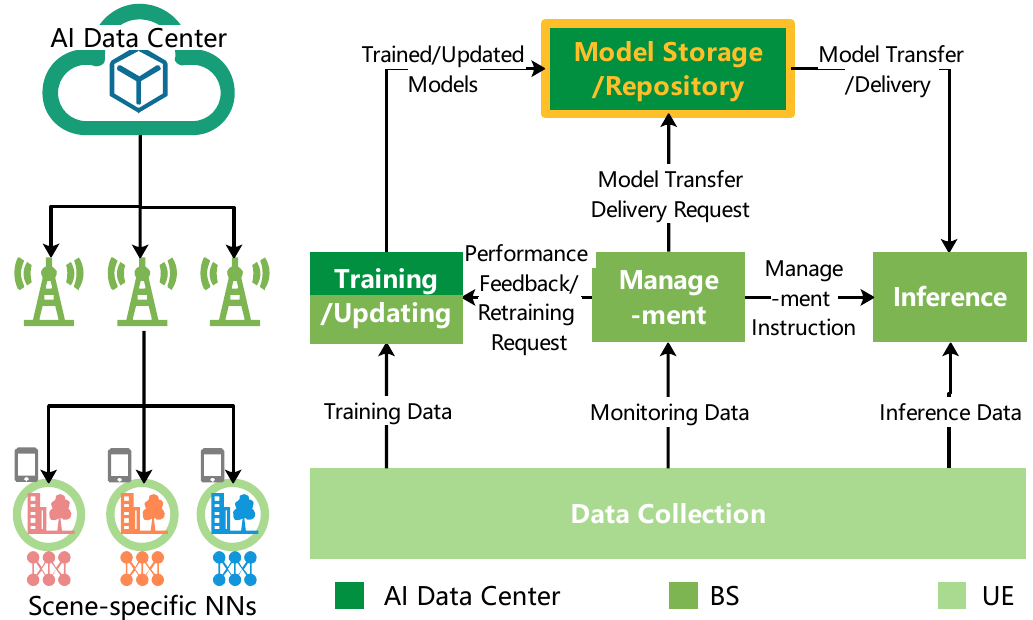}
    \caption{Model repository-based CSI feedback deployment, where the respective functions of the AI data center, BS and UE are marked in different colors.  }
    \label{fig:three-tier}
    \vspace{-5mm}
\end{figure}

Despite these advances, practical deployment of AI-based CSI feedback remains challenging. Current systems typically centralize model training in AI data centers, which either generate personalized models based on CSI uploaded from BSs or construct unified models trained on aggregated data \cite{guo2025prompt, alikhani2024large,mingkai2024task}. Unified models facilitate broad applicability but may exhibit suboptimal performance in specialized environments, while also incurring high computational costs, long training times, and greater reliance on expert tuning. In contrast, scene-specific small models trained on localized CSI can achieve near-optimal performance in their target scenarios, supporting dynamic model switching or rapid fine-tuning during deployment \cite{Li2024Channel,Inoue2023Evalua, Sun2023Int,
Zhang2024Zone-Specific
}. 
For example, the 3GPP R18 \emph{Model Switch} strategy introduces multiple predefined configurations (e.g., 38.901 UMi/UMa) for online selection \cite{3GPP38843,Jing2025Source}. However, models trained for personalized deployment are typically discarded after use, resulting in wasted computational resources and storage. 
More importantly, existing literature predominantly focuses on model training and inference, with limited attention to systematic model storage, management, and reuse across deployments, which constitutes a critical gap that this work aims to address.

To address these limitations, we propose a model repository-based deployment framework, illustrated in Fig.~\ref{fig:three-tier}, which aligns with the 3GPP AI/ML lifecycle management framework~\cite{huang2025aiml}. Our framework focuses primarily on model storage and repository, designing its effective connection with other modules. A centralized repository maintains a catalog of scene-specific AI models, each annotated with standardized metadata. In management, when a BS submits a Delivery Request (retrieval request) by providing a small amount of local CSI or a brief scene description, the repository returns the most appropriate pre-trained model with model transfer for the BS's inference and further model updating, minimizing the need for local fine-tuning and avoiding redundant training.

Efficient model retrieval requires informative yet \textcolor{black}{privacy-aware} metadata. To this end, we adopt and tailor the Learnware paradigm to CSI feedback deployment. Learnware is defined as Model plus Specification~\cite{zhou2016learnware, zhang2021towards, zhou2024learnware, tan2024beimingwu}, where each model is accompanied by a specification that characterizes its capability without exposing the training data.
In this work, we extend the Learnware specification framework to the CSI compression task. Each specification consists of two components. The semantic specification encodes architectural attributes such as input and output dimensions, compression ratios, and task types. The statistical specification summarizes the training data distribution through compact embeddings derived from codebook fingerprint frequency vectors. This design enables efficient model comparison and retrieval without sharing raw CSI.
It has been shown in \cite{lei2024ability} that the specification mechanism in \cite{zhou2016learnware, zhou2024learnware} protects the original training data even against inference and linkage attacks under strong adversarial settings. Therefore, by uploading only statistical specifications rather than raw CSI, BSs \textcolor{black}{avoid exposing raw channel data, mitigating a primary privacy risk,} while enabling rapid model retrieval with low communication overhead.
The main contributions of this work are summarized as follows:
\begin{itemize}

    \item \textbf{First repository-centric CSI feedback deployment architecture.} We propose a principled model repository-based deployment framework for AI-driven CSI feedback, aligned with the 3GPP AI and ML lifecycle management architecture. Unlike conventional retrain-and-discard paradigms, the proposed framework enables systematic model reuse and adaptation across heterogeneous deployment scenarios. This architecture fundamentally transforms CSI feedback from isolated model training to lifecycle-aware model orchestration, substantially reducing redundant training cost and accelerating deployment.

    \item \textbf{Learnware-enabled \textcolor{black}{privacy-enhancing} CSI model indexing.} We introduce a Learnware-based repository design tailored to CSI compression tasks and develop a novel codebook fingerprint $\mathbf{pdf}$ as the statistical specification. The proposed representation captures angular-domain channel structure through DFT beamspace projection while avoiding raw CSI exposure. This design establishes a unified, \textcolor{black}{privacy-enhancing}, and computation-efficient model indexing mechanism that enables scalable model comparison across diverse radio environments.

    \item \textbf{Scalable sub-millisecond retrieval with controllable accuracy-latency tradeoff.} We develop a high-precision and low-latency model retrieval framework based on $\mathbf{pdf}$ similarity matching, achieving over 90\% selection accuracy. To meet stringent CSI feedback timing constraints, we further design a multi-level locality-sensitive hashing (LSH) structure that scales to billion-model repositories and enables 0.25--1.15 ms retrieval latency. The proposed mechanism provides a controllable tradeoff between retrieval accuracy and latency through adjustable search depth, making it suitable for real-time wireless systems.
    
\end{itemize} 

The remainder of this paper is organized as follows. Section~\ref{system model} describes the system and channel models. Section~\ref{Learngene-framework} introduces the Learnware-based repository structure and the specification designs. Section~\ref{sec:architecture-deployment} displays the online deployment framework with Learnware repository. Section~\ref{Simulation-Results} presents simulation settings, comparative evaluations, and ablation studies. Section~\ref{conclusion} concludes the paper and outlines future work.

\section{System Model}
\label{system model}

\subsection{Massive MIMO FDD System}
\label{MIMO system model1}
Consider a downlink massive MIMO system where a BS equipped with $N_{\mathrm{t}}$ uniform linear array (ULA) antennas serves a UE with $N_{\mathrm{r}}$ antennas. The received signal at the UE is given by \cite{guo2022overview, wen2018deep}:
\begin{equation}
\mathbf{y} = \mathbf{H}\mathbf{v}s + \mathbf{n}, \label{eq:received}
\end{equation}
where $\mathbf{H} \in \mathbb{C}^{N_{\mathrm{r}} \times N_{\mathrm{t}}}$ is the channel matrix, $\mathbf{v} \in \mathbb{C}^{N_{\mathrm{t}}}$ is the precoding vector with $\|\mathbf{v}\|_2 = 1$, $s \in \mathbb{C}$ is the transmitted symbol with $\mathbb{E}[|s|^2] = 1$, and $\mathbf{n} \in \mathbb{C}^{N_{\mathrm{r}}}$ is the additive white Gaussian noise with $\mathbf{n} \sim \mathcal{CN}(0, \sigma^2 \mathbf{I}_{N_{\mathrm{r}}})$.
The channel matrix can be decomposed via singular value decomposition (SVD) as:
\begin{equation}
\mathbf{H} = \mathbf{U}\boldsymbol{\Sigma}\mathbf{V}^{H}, \label{eq:svd}
\end{equation}
where $\mathbf{U} \in \mathbb{C}^{N_{\mathrm{r}} \times N_{\mathrm{r}}}$ and $\mathbf{V} \in \mathbb{C}^{N_{\mathrm{t}} \times N_{\mathrm{t}}}$ are unitary matrices, and $\boldsymbol{\Sigma} \in \mathbb{R}^{N_{\mathrm{r}} \times N_{\mathrm{t}}}$ is a diagonal matrix containing the singular values in descending order. For single-stream transmission, the BS selects the first column of $\mathbf{V}$, corresponding to the largest singular value, as the precoding vector $\mathbf{v}$ ~\cite{chen2022implicit}.

\begin{figure*}[t]
    \centering
    \subfigure[Simulation layout.\label{fig:test_layout}]{\includegraphics[width=0.325\linewidth]{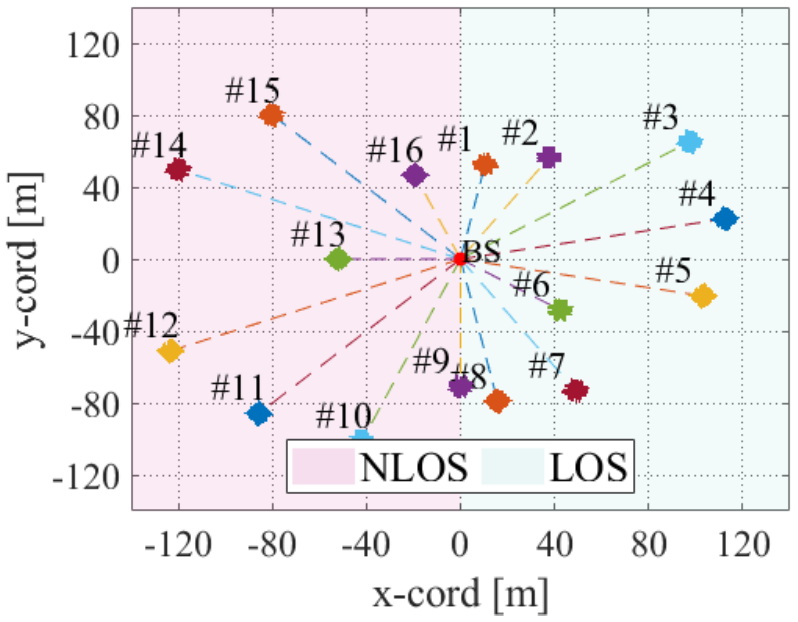}}
    \subfigure[Codeword fingerprint distribution $\mathbf{pdf}$.\label{fig:test_FP}]{\includegraphics[width=0.43\linewidth]{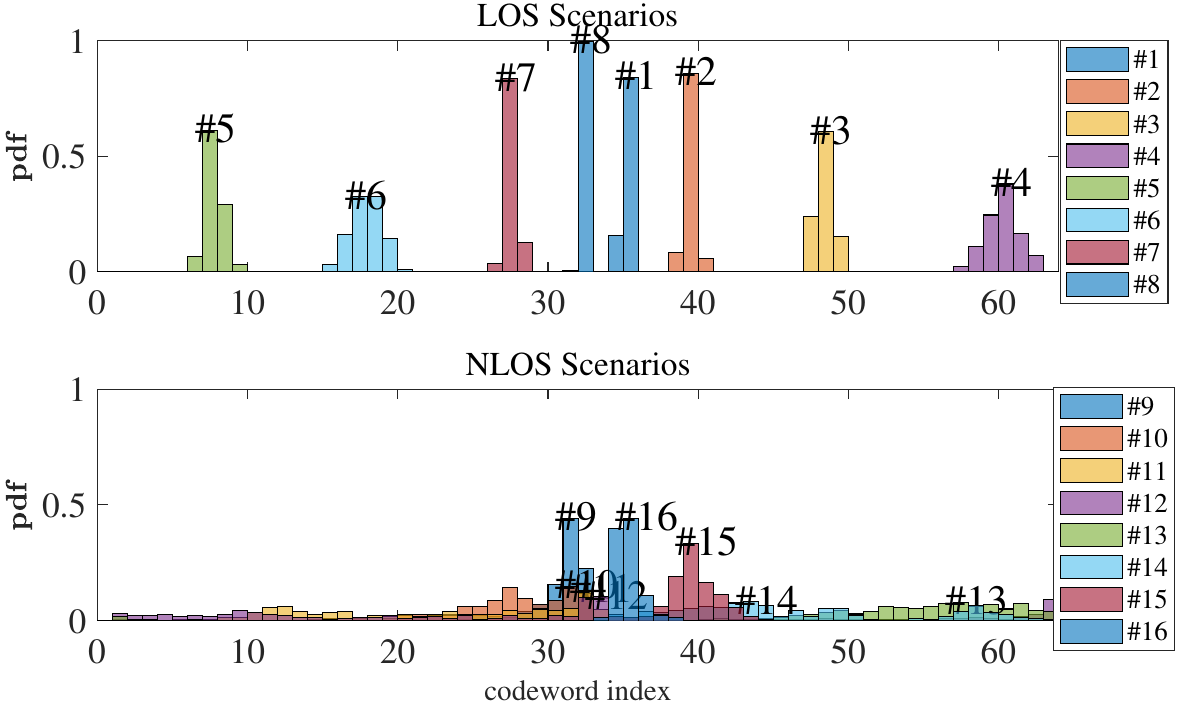}}
    \caption{Sample frequency distribution of codeword fingerprints. (a) Layout of QuaDRiGa-simulated scenarios with UEs randomly distributed in $5\,\rm{m}\times5\,\rm{m}$ rectangle subregions. (b) Corresponding codeword fingerprint distributions $\mathbf{pdf}$ for LOS and NLOS conditions.}
    \label{fig:codeword_FP_layout}
    \vspace{-5mm}
\end{figure*}

\subsection{DL-Based CSI Feedback}

The DL-based CSI feedback framework adopts an autoencoder architecture. The input to the network is the precoding vector $\mathbf{v}$, which is preprocessed by separating its real and imaginary parts and stacking them into a matrix of shape $(2, N_{\mathrm{t}})$. The matrix is then normalized to the range $[0, 1]$ and still denoted as $\mathbf{v}$.

The autoencoder comprises an encoder $\mathsf{enc}(\cdot)$ and a decoder $\mathsf{dec}(\cdot)$, deployed at the UE and BS, respectively. The feedback bitstream is obtained through encoder compression followed by a uniform quantizer $f_{\text{Q}}$ with quantization level $Q$:
\begin{equation}
    \mathbf{s}_{\text{Q}} = f_{\text{Q}}(\mathsf{enc}(\mathbf{v}; \boldsymbol{\Omega}_{\mathrm{enc}})),
\end{equation}
where $\boldsymbol{\Omega}_{\mathrm{enc}}$ denotes the encoder network parameters. 
The $M$-length bitstream $\mathbf{s}_{\text{Q}}$, with compression ratio $\mathrm{CR} = M / (2Q\times N_{\mathrm{t}})$.
At the BS, the decoder reconstructs the precoding vector:
\begin{equation}
\hat{\mathbf{v}} = \mathsf{dec}(f^{-1}_{\text{Q}}(\mathbf{s}_{\text{Q}}); \boldsymbol{\Omega}_{\mathrm{dec}}),
\label{equ:functional}
\end{equation}
where $\boldsymbol{\Omega}_{\mathrm{dec}}$ are the decoder parameters.

To evaluate reconstruction fidelity, we adopt the squared generalized cosine similarity (SGCS) metric:
\begin{equation} 
\rho^2 = \frac{1}{|\mathcal{D}|} \sum_{\mathbf{v} \in \mathcal{D}} \left( \frac{|\mathbf{v}^{H} \hat{\mathbf{v}}|}{\|\mathbf{v}\|_2 \|\hat{\mathbf{v}}\|_2} \right)^2.\label{equ:SGCS}
\end{equation}
During end-to-end training, we minimize the cosine distance between the original and reconstructed precoding vectors:
\begin{equation}
\mathcal{L} = \frac{1}{|\mathcal{D}|} \sum_{\mathbf{v} \in \mathcal{D}} \left(1 - \frac{|\mathbf{v}^H \hat{\mathbf{v}}|}{\|\mathbf{v}\|_2 \|\hat{\mathbf{v}}\|_2}\right),
\label{equ:cosine_loss}
\end{equation}
where $\mathcal{D}$ denotes the training dataset and $|\mathcal{D}|$ its cardinality. This loss function is directly aligned with the SGCS metric, thereby ensuring consistency between the training objective and the evaluation criterion.

\subsection{Codeword Fingerprint}

To characterize the statistical properties of CSI datasets, we adopt a codebook fingerprinting technique based on \cite{baur2024evaluation}. We utilize a discrete Fourier transform (DFT) codebook, which is well-suited to uniform linear arrays and captures long-term angular characteristics of wireless channels \cite{Suh2017Construction}.

The DFT codebook of size $N_{\mathrm{code}} = 2^B$ is defined as $\mathbf{W} = [\mathbf{w}_0, \mathbf{w}_1, \ldots, \mathbf{w}_{\textcolor{black}{N_{\mathrm{code}}-1}}]$, where the $n$-th codeword is given by:
\begin{equation}
\mathbf{w}_n = \frac{1}{\sqrt{N_{\mathrm{t}}}} \left[1, e^{j \frac{2\pi}{N_{\mathrm{code}}} n}, \ldots, e^{j \frac{2\pi}{N_{\mathrm{code}}}(N_{\mathrm{t}}-1) n} \right]^{T}.
\end{equation}
For a given precoding vector $\mathbf{v}$, its fingerprint in the angular domain is computed as:
\begin{equation}   
\mathbf{f} =  \mathbf{v}^{H} \mathbf{W}  = [f_1, f_2, \ldots, f_{N_{\mathrm{code}}}].
\end{equation}  
The index of the dominant beam direction is:
\begin{equation}
\text{idx} = \mathop{\mathrm{argmax}}\limits_{i \in \{1, \ldots, N_{\mathrm{code}}\}} |f_i|.
\end{equation}

By aggregating the dominant directions over the dataset, we compute a histogram to approximate the angular probability density function (PDF):
\begin{equation}
\mathbf{pdf} = \mathrm{hist}_{\mathcal{D}}(\text{idx}).
\label{Equ:pdf}
\end{equation}
This statistical distribution $\mathbf{pdf}$ reflects the angular spread of energy in the dataset, revealing environmental characteristics. In line-of-sight (LOS) cases, the dominant direction corresponds to the angle of departure (AoD) of the LOS path. In non-line-of-sight (NLOS) scenarios, it reflects the AoD of the strongest scattering path.

We illustrate this with CSI datasets generated using QuaDRiGa under the ``3GPP-38.901-UMi-LOS/NLOS'' settings. UEs are randomly distributed within $5\,\rm{m}\times5\,\rm{m}$ rectangle subregions, as shown in Fig.~\ref{fig:test_layout}. The resulting $\mathbf{pdf}$ distributions are presented in Fig.~\ref{fig:test_FP}, where LOS scenarios exhibit more concentrated angular energy, while NLOS cases show broader dispersion due to multipath scattering. 
The LOS case illustrates an extreme example where small sampling areas ($5\,$m) and ULA configuration create concentrated angular distributions. This apparent correlation is an artifact of experimental design: expanding sampling areas yields diffuse distributions, different antenna arrays break the one-to-one mapping, and each dataset restarts obstacle layouts across independent maps. Unlike CSI-positioning with fixed environments, our framework prevents \textcolor{black}{inferring the} precise UE position from $\mathbf{pdf}$.

\section{Learnware Framework}\label{Learngene-framework}

This section presents our Learnware‐based CSI‐feedback model repository. We first motivate the use of scene‐specific small models and the integration of a centralized repository into the CSI‐feedback workflow. We then outline the Learnware paradigm and its benefits for CSI‐feedback applications. 

\subsection{Motivation}
Deploying AI‐driven CSI‐feedback models entails two primary challenges: (1) balancing generalization with per‐scene performance, and (2) managing the computational and deployment costs of large models. Our Learnware Model Repository addresses these by enabling efficient reuse of scene‐specific models identified via statistical specifications. We highlight three key insights: the efficacy of small models, the power of statistical specifications, and the accelerated fine‐tuning convergence they afford.

\subsubsection{Scene‐Specific Small Models vs.\ General‐Purpose Models}
Training a single large‐capacity model to generalize across diverse scenarios simplifies management but demands extensive compute, memory, and often yields suboptimal performance in specific environments. In contrast, scene‐specific small models, trained on localized CSI data, capture fine‐grained propagation characteristics, require fewer resources, and consistently outperform general models within their target domains (see Table~\ref{tab:generalization}). Detailed experiments appear in Section~\ref{Simulation-Results}. This phenomenon is theoretically grounded in multi-task learning theory~\cite{baxter2000model, maurer2016benefit}. A general model trained on heterogeneous scenes suffers from negative transfer\textcolor{black}{:} parameter sharing induces gradient interference across tasks, degrading performance when tasks are insufficiently related. Scene-specific models avoid this interference by dedicating capacity to a single distribution, yielding superior performance under capacity-limited regimes.

\subsubsection{Statistical Specifications vs.\ Semantic Specifications}
Accurate model retrieval requires more than coarse structural descriptors (e.g., network depth, codeword length). Semantic specifications alone cannot distinguish differences in data distributions due to environmental variations. Instead, we use \emph{statistical specifications}, continuous descriptors of the training‐data distribution (e.g., codeword fingerprint histograms), to index models in a continuous specification space. This approach generalizes to unseen environments and avoids the inefficiency of exhaustive validation.

\subsubsection{Rapid Fine‐Tuning Convergence}
\begin{figure}
    \centering
    \includegraphics[width=0.67\linewidth]{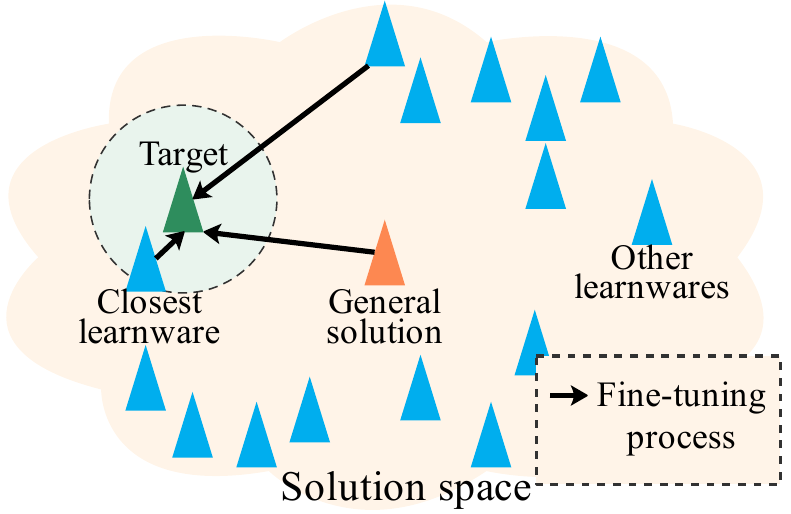}
    \caption{CSI feedback model solution space. Curves indicate convergence paths from (1) a mismatched model, (2) a general model, and (3) the closest Learnware model.}
    \label{fig:solution-space}
    \vspace{-6mm}
\end{figure}

Retrieved models often require local fine‐tuning. The closer a model’s original training distribution is to the target domain, the fewer iterations and samples needed for convergence \cite{dwivedi2019representation}. Fig.~\ref{fig:solution-space} illustrates convergence trajectories from (1) a mismatched model, (2) a general model, and (3) the nearest Learnware model. A densely populated repository ensures rapid adaptation or even direct deployment without further training.

\subsection{Learnware Paradigm and Specification Design}
\label{subsec:Learnware paradigm}

\begin{figure*}
    \centering
    \includegraphics[width=0.9\linewidth]{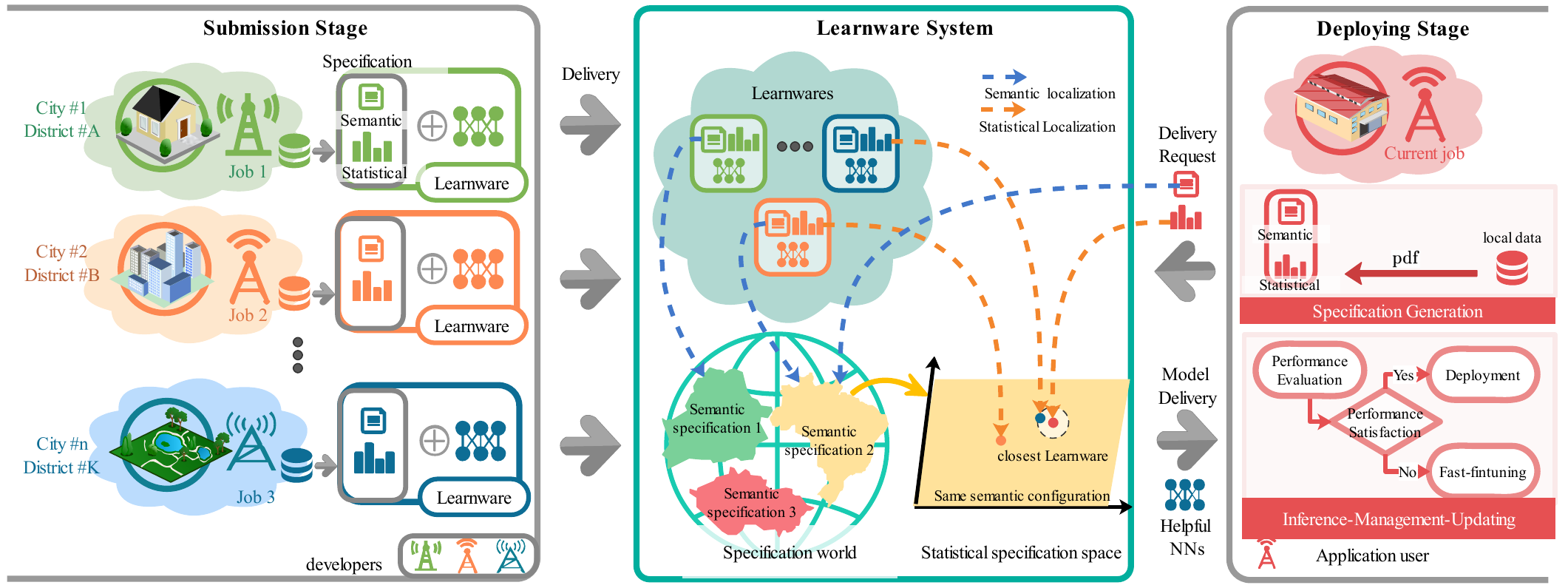}
    \caption{Learnware paradigm applied to CSI feedback neural networks.}
    \label{fig:Learnware-paradigm}
    \vspace{-4mm}
\end{figure*}

The motivations described in the previous subsection support our design of a Learnware-based repository that emphasizes scene-specific model reuse, retrieval using statistical specifications, and efficient adaptation through fine-tuning.
Fig.~\ref{fig:Learnware-paradigm} illustrates the framework of the Learnware-based repository, which consists mainly of two stages: \emph{submission} and \emph{deployment}. In the submission stage, developers package each model with its semantic and statistical specifications and upload it to the repository. In the deployment stage, users submit their task requirements, and the system matches semantic tags and locates the appropriate \emph{specification} for model retrieval. 

The Learnware paradigm \cite{zhou2016learnware, zhou2024learnware} provides a marketplace of pre-trained models that users can reuse without sharing raw data. Each Learnware entry contains a model and its specification, allowing the system to match user requests \textcolor{black}{while keeping each contributor's training data local}. 
This paradigm is provided as a heterogeneous and evolving alternative to the traditional approach of training strong generalized models from aggregated datasets without collecting their training data. Instead of learning from a large centralized dataset~\cite{guo2025large}, it accumulates knowledge through an ever-growing collection of distributed models. This model-centric approach supports on-demand access to task-relevant knowledge while avoiding catastrophic forgetting and \textcolor{black}{the scalability limitations of} data aggregation, \textcolor{black}{as} depicted in Tab.~\ref{tab:paradigm_comparison}. It also offers the \textcolor{black}{future possibility of} reassembling multiple models to address complicated or even unplanned \textcolor{black}{tasks}~\cite{zhou2016learnware}.

\begin{table}[t]
\centering
\caption{Comparison of Paradigms}
\resizebox{0.49\textwidth}{!}{%
    \renewcommand\arraystretch{0.8}
\begin{tabular}{c|ccccc}
\toprule
\multirow{2}{*}{\textbf{Paradigm}} & \multirow{2}{*}{\begin{tabular}[c]{@{}c@{}} \textbf{Resource} \\ \textbf{requirements} \end{tabular}}
& \multirow{2}{*}{\begin{tabular}[c]{@{}c@{}} \textbf{Transmission} \\ \textbf{cost} \end{tabular}}
& \multirow{2}{*}{\begin{tabular}[c]{@{}c@{}} \textbf{Lifelong} \\ \textbf{learning} \end{tabular}}
& \multirow{2}{*}{\begin{tabular}[c]{@{}c@{}} \textbf{Robustness} \\ \textbf{to forgetting} \end{tabular}}
& \multirow{2}{*}{\begin{tabular}[c]{@{}c@{}} \textbf{Data} \\ \textbf{privacy} \end{tabular}}
\\
& & & & &\\
\midrule
\textbf{Aggregate datasets} & High & High & $\times$ & $\times$ & $\times$ \\
\textbf{Learnware Paradigm} & Low & Low & \checkmark & \checkmark & \checkmark \\
\bottomrule
\end{tabular}}
\label{tab:paradigm_comparison}\vspace{-3mm}
\end{table}
The remainder of this section describes the submission stage and the design of specifications. The deployment stage will be discussed in the next section.

\subsubsection{Model Submission}

The Learnware system supports large-scale operation, allowing many developers to contribute models, thus creating a rich, diverse repository. 
Future users submit requirements to the Learnware Dock system, which identifies or reassembles learnwares. Users can then reuse these models or refine them with their data.
Although BSs are encouraged to contribute high-quality models voluntarily, most CSI-feedback models are trained centrally in an AI resource data center, which can maintain a broad collection of pre-trained models. Each model is stored as a Learnware entry that includes its specifications, describing functional characteristics, enabling efficient identification and retrieval without retaining the original training data.

\subsubsection{Specification Design}

Due to the diverse functionalities and training data distributions of Learnware models, the specification space is inherently complex and expansive. Without a well-structured design, users may struggle to locate suitable models efficiently. To address this, we introduce the concept of a \emph{specification world}, where functionally similar models are grouped into \emph{specification islands}. Within each island, models differ primarily in their training data distributions, referred to as the data statistical space.

Thus, the overall specification world can be viewed as a coupling of a discrete functional space and a continuous statistical space. For two models with identical functional specifications, a smaller difference in their training data distributions generally implies better generalization performance when transferred to similar environments. As shown in Fig.~\ref{fig:Learnware-paradigm}, users first identify the appropriate island based on semantic requirements, and then select the closest model by comparing statistical similarities between their local data and the Learnware training distribution.

\begin{figure*}[t]
\centering
\subfigure[Euclidean distance matrix of 144 datasets. \label{fig:Distance-performance1}]{\includegraphics[width=0.32\linewidth]{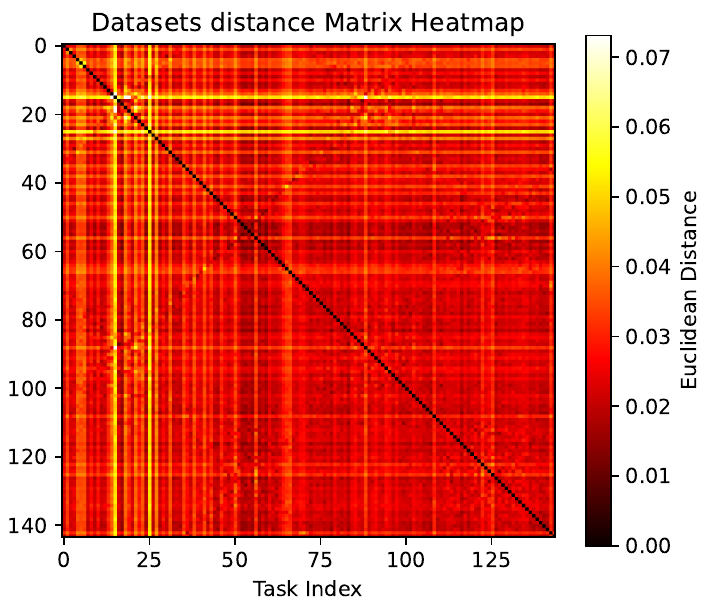}}
\subfigure[Cosine similarity matrix of $\mathbf{pdf}$. \label{fig:Distance-performance2}]{\includegraphics[width=0.32\linewidth]{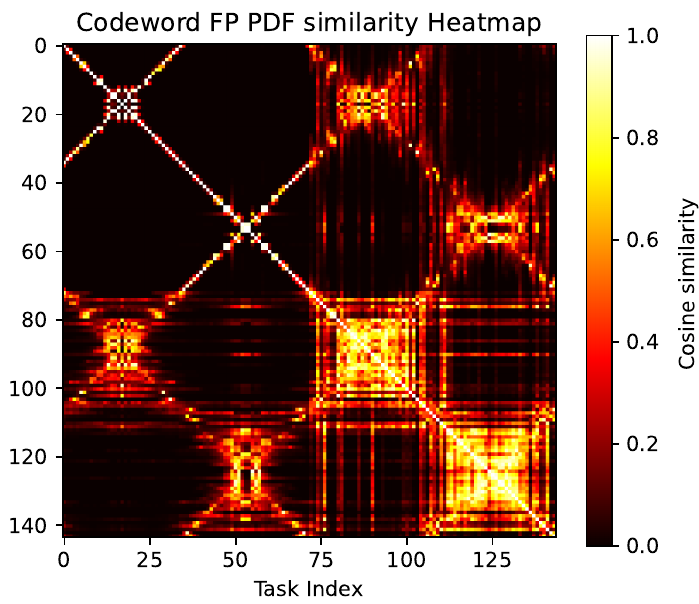}}
\subfigure[Performance evaluation matrix. \label{fig:Distance-performance3}]{\includegraphics[width=0.32\linewidth]{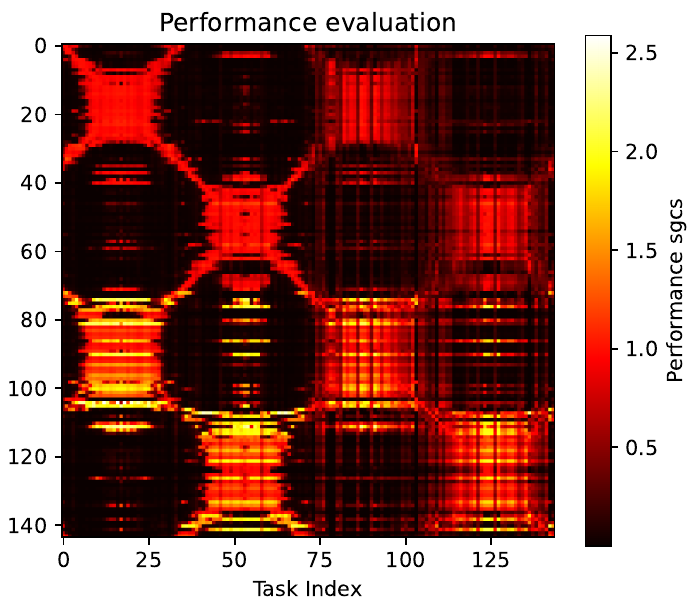}}
\caption{Comparison between dataset-level distances and model generalization performance.}
\label{fig:Distance-performance}
\vspace{-4mm}
\end{figure*}
 
Specifications are divided into two categories:
\paragraph{Semantic Specifications} These consist of string-based descriptors defining model functionality. They help identify the relevant specification island and cover attributes such as task type (e.g., full or implicit feedback, one-side or two-side network), model structure (input/output dimensions, codeword length), and system configurations (bandwidth, frequency, number of carriers, antenna array type). An illustrative example is provided in Tab.~\ref{tab:Semantic-specification}.

\paragraph{Statistical Specifications} These characterize the distributional properties of the training data and help pinpoint a model's exact position within an island. Due to privacy concerns, the original training data are not shared. Instead, statistical specifications are generated offline at the model developer's side, and packaged with the model and semantic tags before uploading. In our framework, the codeword fingerprint distribution $\mathbf{pdf}$, defined in Eq.~\ref{Equ:pdf}, is used as the statistical specification. Cosine similarity is employed as the distance metric for retrieval.

\subsubsection{Validation of Codeword Fingerprint PDF}
We validate the effectiveness of the codeword fingerprint distribution $\mathbf{pdf}$ as a statistical specification. Following the dataset similarity evaluation framework established in~\cite{Morais2024DatasetSimilarity}, we investigate the correlation between dataset similarity and model generalization performance. The underlying principle is that if a distance metric meaningfully captures dataset characteristics, then datasets that are close in that metric space should yield similar model performance when models are transferred between them. This validation is essential for our Learnware retrieval mechanism, which relies on $\mathbf{pdf}$ similarity to select appropriate pre-trained models.

To this end, we use 144 scenario-specific simulated datasets (i.e., Learnwares), comprising 72 LOS and 72 NLOS environments. These datasets are arranged counterclockwise according to the azimuth angle relative to a central BS, ensuring full 360-degree coverage for both propagation conditions.
Fig.~\ref{fig:Distance-performance1} presents the Euclidean distance matrix of raw CSI vectors between dataset pairs. Each element $(i,j)$ is calculated as:
\begin{equation}
d_{(i,j)} = \frac{1}{|\mathcal{D}_i| \times |\mathcal{D}_j|} \sum_{\mathbf{v}^{(i)} \in \mathcal{D}_i} \sum_{\mathbf{v}^{(j)} \in \mathcal{D}_j} \left\| \mathbf{v}^{(i)} - \mathbf{v}^{(j)} \right\|_2.
\end{equation}
Fig.~\ref{fig:Distance-performance2} shows the cosine similarity matrix between the $\mathbf{pdf}$ vectors of the datasets, where each element is given by:
\begin{equation}
{\rm sim}_{(i,j)} = \frac{|\mathbf{pdf}_{i}^H \mathbf{pdf}_{j}|}{\|\mathbf{pdf}_i\|_2  \|\mathbf{pdf}_j\|_2}.
\end{equation}
Fig.~\ref{fig:Distance-performance3} displays the generalization performance matrix. The element $p_{(i,j)}$ indicates the SGCS performance of a model trained on dataset $\mathcal{D}_i$ and tested on $\mathcal{D}_j$. To ensure fairness, all values are normalized by the self-performance $p_{(j,j)}$.

A comparison of the three matrices reveals that Fig.~\ref{fig:Distance-performance2} aligns more closely with Fig.~\ref{fig:Distance-performance3} than does Fig.~\ref{fig:Distance-performance1}, suggesting that $\mathbf{pdf}$ is a more effective proxy for predicting model transferability than raw CSI distances. This confirms the suitability of $\mathbf{pdf}$ as a statistical specification for Learnware retrieval.
Notably, the cosine similarity matrix in Fig.~\ref{fig:Distance-performance2} exhibits greater sharpness than the performance matrix in Fig.~\ref{fig:Distance-performance3}. For example, the X-shaped spot in the upper-left corner of Fig.~\ref{fig:Distance-performance3} is more diffuse than its counterpart in Fig.~\ref{fig:Distance-performance2}, indicating that $\mathbf{pdf}$ similarities serve as a reliable surrogate for functionality proximity. This X-pattern arises from the simulation configuration, in which the BS is a north-south ULA and datasets are spatially distributed by angular sectors, yielding a periodic structure in beam directionality.

A closer analysis comparing LOS and NLOS subsets reveals that $\mathbf{pdf}$ is more sensitive to variations in LOS scenes, as evidenced by the different sharpness of the upper-left submatrix (LOS) and the lower-right submatrix (NLOS) in Fig.~\ref{fig:Distance-performance2}. This is expected since $\mathbf{pdf}$ in LOS conditions mainly captures the AoD of the direct path, which is sharply defined by the relative geometry between the UE and BS. In contrast, $\mathbf{pdf}$ in NLOS conditions reflects a broader scattering profile due to multipath propagation, resulting in a more diffuse distribution.

In the early stages of repository development, the diversity of available models may be limited. To mitigate this, initial configuration sets can be used to define semantic specification labels, thereby establishing a range of specification islands. Moreover, simulation-trained CSI models can be employed to populate the repository. As more models are added over time, both semantic and statistical specification coverage will naturally expand.

\begin{figure*}[t]
    \centering
    \setlength{\abovecaptionskip}{-1.5mm}
    \setlength{\belowcaptionskip}{-3cm}
    \subfigure[Scheme with Original model repository. \label{fig:online-deployment1}]{\includegraphics[width=0.38\linewidth]{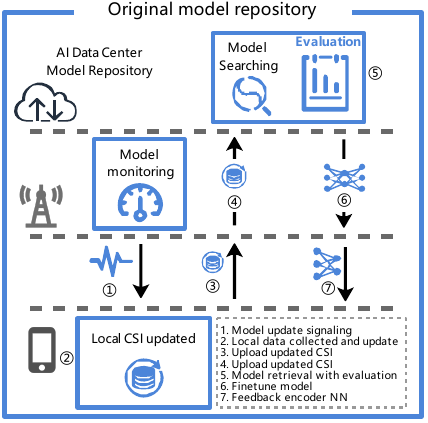}}
    \subfigure[Scheme with Learnware model repository.\label{fig:online-deployment2}]{\includegraphics[width=0.38\linewidth]{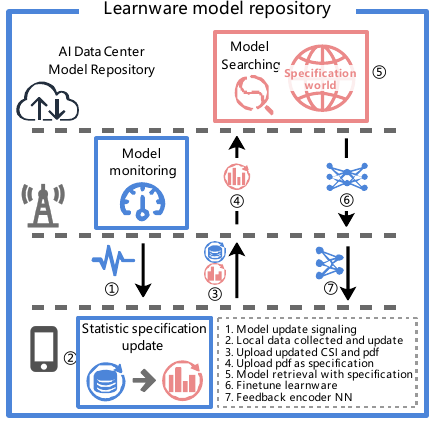}}
    \caption{Online deployment schemes based on model repository, where the key differences are marked in orange.}
    \label{fig:deplyment}
    \vspace{-5mm}
\end{figure*}

\section{Deployment Strategies}
\label{sec:architecture-deployment}

This section describes the deployment stage of our Learnware framework, including the online retrieval mechanism, comparisons with alternative methods, and an optimized search strategy to accelerate model lookup.

\subsection{Deploying Stage}

During deployment, the user-side BS first activates Model Management by monitoring for model mismatch. It then submits a model Delivery Request to the Learnware marketplace, which identifies and delivers the most suitable models. This process presents two main challenges: (1) effectively retrieving models that satisfy the user's task requirements; and (2) maximizing the utility of the returned models while minimizing adaptation overhead.

The baseline approach, denoted the Original Model Repository, uploads a subset of local CSI data as a validation set, evaluates every candidate model on this set, and returns the highest‐performing model. Although this method achieves high retrieval accuracy, it suffers from several drawbacks: it compromises data privacy and intellectual property, incurs substantial communication overhead, and scales poorly due to sequential model evaluations. These limitations render it impractical for real‐time CSI deployment.

To overcome these issues, our Learnware system employs a specification space retrieval mechanism, as illustrated in Fig.~\ref{fig:Learnware-paradigm}. Local CSI data is only used in specification generation, and the specification is sent to the AI data center as the request. The semantic task specifications are used to locate the appropriate specification island. Within that island, statistical specifications are used to index a suitably similar model. Crucially, an exact statistical match is not required. As illustrated in Fig.~\ref{fig:Distance-performance}, small discrepancies often result in highly interoperable models, enabling flexible and responsive retrieval.
When multiple candidates share identical statistical specifications but differ in network architecture, the BS performs a lightweight on-device evaluation. If the returned model meets the performance threshold, it is deployed directly; otherwise, the BS fine-tunes the model locally.

In the early stages of repository development, coverage of the specification space may be sparse, leading to suboptimal matches. Moreover, the BS may lack knowledge of the ideal local performance, making it difficult to assess the gap between the retrieved model and an optimally trained one. We address this by incorporating a fine-tuning step governed by a user-defined threshold, based on available samples, allowable training iterations, and resource constraints. Models that fail to improve after fine-tuning are discarded in favor of the original, while those that show improvement are adopted for deployment. 

\begin{table*}[t]
    \setlength{\abovecaptionskip}{0cm}
    \setlength{\belowcaptionskip}{-0.2cm}
    \renewcommand\arraystretch{1}
    \centering
    \caption{\label{tab:Advantages}Comparison of general model, model switch, and model repository-based schemes across key deployment considerations.}

	\resizebox{0.98\textwidth}{!}{
\begin{tabular}{c c ccccc} \toprule
\multicolumn{2}{c}{\textbf{Method}}   & \textbf{General Model} & \textbf{General Model with fine-tuning}  & \textbf{Model Switch} & \textbf{Original Model Repository} & \textbf{Learnware Model Repository}  \\ \midrule
\multirow{2}{*}{\begin{tabular}[c]{@{}c@{}}\textbf{Construction}\\ \textbf{(Initialize)}\end{tabular}} & AI data center & \multicolumn{2}{c}{Train one model using large-scale data}  &\begin{tabular}[c]{@{}c@{}}Train several models \\ using large-scale data \end{tabular} & \multicolumn{2}{c}{Train many models in parallel}  \\
 & BS  & \multicolumn{2}{c}{Store a single model}    & Store several models   & \multicolumn{2}{c}{Store a single model} \\ \midrule
\multirow{3}{*}{\begin{tabular}[c]{@{}c@{}} \textbf{Online} \\ \textbf{deployment} \end{tabular}} & AI data center & None & None & None & \multicolumn{2}{c}{Model retrieval and return}  \\
   & BS   &  None & \multicolumn{1}{c}{\begin{tabular}[c]{@{}c@{}}Fine-tune and \\ transmit encoder NN\end{tabular}} & \multicolumn{1}{c}{\begin{tabular}[c]{@{}c@{}}Switch model and \\ transmit encoder NN\end{tabular}}       & \multicolumn{1}{c}{\begin{tabular}[c]{@{}c@{}}Upload local CSI samples, \\fine-tune, transmit encoder NN \end{tabular}} & \multicolumn{1}{c}{\begin{tabular}[c]{@{}c@{}}Upload specifications, \\fine-tune, transmit encoder NN \end{tabular}} \\
& UE & None & \multicolumn{4}{c}{Upload local CSI samples to BS}  \\ \midrule
\multirow{4}{*}{\begin{tabular}[c]{@{}c@{}} \textbf{Merits} \\ \textbf{and} \\ \textbf{Limitations} \end{tabular}}   & Accuracy   & Low  & High   & Middle  & High   & High  \\
& Communication overhead & Zero & High  & Low   & Middle  & Middle \\
& Computation overhead   & Zero & High  & Low   & High    & Middle \\
& Data privacy           &\Checkmark   & \Checkmark   & \Checkmark & \XSolidBrush &  \Checkmark          \\ \bottomrule                                                                                                    
\end{tabular}}
\vspace{-5mm}
\end{table*}

\subsection{Post-Retrieval Adaptation}

Online deployment of CSI feedback models involves interactions among the AI data center, the BS, and the UE. Initially, the AI data center trains a two-sided neural network and deploys it to the BS for local adaptation. As illustrated in Fig.~\ref{fig:deplyment}, when the BS detects performance degradation, it issues a model update request \cite{Ahmed2018Explicit}. The UE collects new CSI samples and forwards them to the BS, which may either fine-tune its current model, incurring substantial computation and data requirements, or request a new pre-trained model whose training distribution more closely matches the current channel conditions.

As illustrated in Fig.\ref{fig:online-deployment1}, under the Original Model Repository, the BS must upload raw CSI data for exhaustive candidate evaluation, raising privacy and intellectual property concerns and introducing high latency. In contrast, the Learnware Model Repository requires only compact statistical specifications, \textcolor{black}{reducing raw CSI exposure} and significantly reducing retrieval time. Although this approach may slightly reduce selection accuracy, it enables near real-time updates. As illustrated in Fig.\ref{fig:online-deployment2}, semantic specifications (e.g., antenna configuration and bandwidth) remain fixed during deployment, constraining the search to a single specification island. The BS then submits only the relevant statistical specification for rapid retrieval. For large repositories, the elastic search strategies presented in Section~\ref{subsection:search-srategy} maintain fast and scalable performance.

\subsection{Comparison with Alternative Methods}

We evaluate several baselines against our repository-based solutions:
\begin{itemize}
    \item \textbf{General Model:} A single model trained on a large, diverse dataset and deployed directly to the BS.
    \item \textbf{General Model with fine‐tuning:} The general model is fine‐tuned at the BS to match local CSI conditions.
    \item \textbf{Model Switch:} Multiple pre‐trained models with different configurations; the BS selects the best based on local performance.
    \item \textbf{Original Model Repository:} The BS uploads local data to the AI data center, which evaluates all candidates and returns the best for fine‐tuning.
    \item \textbf{Learnware Model Repository:} The BS uploads statistical specifications; the AI data center retrieves the most appropriate pre‐trained model based on semantic and statistical matching.
\end{itemize}

Table~\ref{tab:Advantages} summarizes these approaches across initialization and deployment phases, as well as trade‐offs in accuracy, communication cost, computation overhead, and data privacy. During initialization, all methods require substantial training: the General Model and Model Switch focus on fewer, larger models, whereas repository approaches leverage many smaller models for parallelism and scalability. In deployment, General Model and Model Switch require no further AI data center interaction, while repository methods involve continuous communication. The Learnware Model Repository achieves a balanced trade‐off, delivering high performance with moderate overhead and preserving user privacy, an equilibrium further validated by our experimental results.

\subsection{Adjustable-Speed Model Search Strategy}
\label{subsection:search-srategy}
To support real-time CSI feedback deployment, we enhance the repository's internal structure to enable rapid and scalable retrieval. After locating the appropriate specification island using semantic attributes, brute-force computation of cosine similarities across all candidates remains impractical for large repositories. As demonstrated in Fig.~\ref{fig:Distance-performance}, models with similar statistical specifications tend to be functionally interoperable, offering opportunities to accelerate retrieval. Therefore, we propose an adjustable-speed multi-level LSH strategy (Algorithm~\ref{alg:lsh}), which unifies coarse-grained anchor selection and hierarchical indexing into a $K$-level cascade with early termination capability.

\begin{algorithm}[t]
\small
\caption{Hierarchical LSH Retrieval}
\label{alg:lsh}
\LinesNumbered
\SetAlgoNlRelativeSize{-1}
\KwIn{Repository $\mathcal{M}=\{\mathbf{p}_i\}_{i=1}^N$ ($\mathbf{p}_i$: normalized spec.); query $\mathbf{q}$; levels $K$; LSH params $\{(k_\ell,L_\ell)\}_{\ell=1}^K$ ($k_\ell$: \#hash bits, $L_\ell$: \#tables); stop level $k_{\text{stop}}$}
\KwOut{Retrieved anchor model $\mathbf{m}^*\in\mathcal{M}$}

\tcp*[f]{Offline: build $K$-level index. $\mathbf{W}_\ell\in\mathbb{R}^{L_\ell\times k_\ell\times d}$: random projection matrix}\;
\For{$\ell=1,\ldots,K$}{
    Sample $\mathbf{W}_\ell$ with normalized rows\;
    \lFor{$\mathbf{p}_i\in\mathcal{M}_\ell$, $j=1,\ldots,L_\ell$}{$\mathcal{T}_\ell^{(j)}[h_i^{(j)}]\leftarrow\mathcal{T}_\ell^{(j)}[h_i^{(j)}]\cup\{i\}$ with $h_i^{(j)}=\text{PackBits}(\text{sign}(\mathbf{W}_\ell^{(j)}\mathbf{p}_i))$}
    \lFor{each bucket $h$ with models $\mathcal{B}$}{$\mathcal{A}_\ell^{(j)}[h]\leftarrow\arg\max_{\mathbf{p}\in\mathcal{B}}\mathbf{p}^\top\mathbf{c}$: select model closest to centroid $\mathbf{c}=\frac{1}{|\mathcal{B}|}\sum_{\mathbf{p}'\in\mathcal{B}}\mathbf{p}'$}
}

\tcp*[f]{Online: cascade. $h^{(j)}=\text{PackBits}(\text{sign}(\mathbf{W}_\ell^{(j)}\mathbf{q}))$: query signature}\;
\For{$\ell=1,\ldots,k_{\text{stop}}$}{
    $\mathbf{m}_\ell\leftarrow\arg\max_{j=1,\ldots,L_\ell}\mathbf{q}^\top\mathcal{A}_\ell^{(j)}[h^{(j)}]$ \tcp*{best anchor at level $\ell$}
    \lIf{$\mathbf{q}^\top\mathbf{m}_\ell>\theta_\ell$ (confidence threshold)}{return $\mathbf{m}_\ell$}
}
\Return{$\mathbf{m}^*\leftarrow\mathbf{m}_{k_{\text{stop}}}$}\;
\end{algorithm}

The algorithm consists of two phases. \textbf{Offline indexing} builds $K$ levels of LSH tables with increasing hash precision. For each level $\ell$, we generate $L_\ell$ hash tables using $k_\ell$-bit signatures ($k_1 < k_2 < \cdots < k_K$, $L_1 > L_2 > \cdots > L_K$). This configuration ensures that coarse levels use more tables for high recall to capture a broad candidate set, while finer levels use fewer tables with longer hash keys for precise filtering, balancing retrieval efficiency and accuracy. Each bucket pre-computes an anchor model selected as the one closest to the centroid, enabling $O(1)$ retrieval without enumerating all candidates. \textbf{Online retrieval}  processes levels sequentially until a specified stopping level $k_{\text{stop}}$, which can be determined by a latency budget $\tau$ or confidence thresholds $\{\theta_\ell\}$. At each level, the query specification $\mathbf{q}$ is hashed to retrieve $L_\ell$ bucket anchors, and the best anchor $\mathbf{m}_\ell$ is selected via cosine similarity. The system can terminate early if $\mathbf{q}^T\mathbf{m}_\ell > \theta_\ell$, or proceed to finer levels for higher accuracy.

This design provides adjustable speed through two mechanisms: (i) \textit{level selection}: stopping at coarse levels (small $k_{\text{stop}}$) yields sub-millisecond latency suitable for urgent updates, while full cascade (large $k_{\text{stop}}$) ensures near-optimal accuracy; (ii) \textit{confidence-based early exit}: termination triggers when retrieved anchors already satisfy performance requirements, avoiding unnecessary computation. The per-level complexity is $O(L_\ell k_\ell d + L_\ell d)$, independent of repository size $N$, enabling scalable deployment to billion-model repositories.

\section{Simulation Results and Discussions}
\label{Simulation-Results}

\subsection{Learnware Model Repository Initiation}

We initialized the Learnware model library using the implicit feedback CSI two-sided network as an example to construct a specification island. The corresponding semantic labels are shown in Tab.~\ref{tab:Semantic-specification}.

\begin{table}[t]
    \setlength{\abovecaptionskip}{0cm}
    \setlength{\belowcaptionskip}{-0.2cm}
    \centering
    \caption{\label{tab:Semantic-specification}Semantic specification labels used in the experiment.}
    \renewcommand\arraystretch{1}
    \resizebox{0.3\textwidth}{!}{
    \begin{tabular}{|l|l|}
        \hline
        \texttt{CSI\_Number\_Domain} & {spatial domain} \\ \hline
        \texttt{Feedback\_Mode} & {implicit feedback} \\ \hline
        \texttt{Structure} & {two-sided model} \\ \hline
        \texttt{Input\_Dimension} &  {$32\times 2$} \\ \hline
        \texttt{Output\_Dimension} & {$32\times 2$} \\ \hline
        \texttt{Compression\_Ratio} & {1/16} \\ \hline
        \texttt{Quantization\_Bits} & {4} \\ \hline
        \texttt{BS\_Antenna\_Array} & {ULA} \\ \hline
        \texttt{Central\_Frequency} & {2.6 GHz} \\ \hline
        \texttt{Bandwidth} & {20 MHz} \\
        \hline
    \end{tabular}}
\vspace{-0.3cm}
\end{table}

\subsubsection{CSI Simulation}
Before collecting models, we generated simulation datasets using the configurations listed above to train Learnware models and enrich the specification island. 
We created 144 Learnwares using QuaDRiGa~\cite{QuaDRiGa} under the 3GPP-38.901-UMi configuration (72 LOS + 72 NLOS, each with 10,000 samples). CSI sampling covered full 360$^\circ$ azimuth diversity with independent environment layouts per dataset (i.e., the environment layout was restarted for each dataset). Details are in Tab.~\ref{tab:QuaDRiGaSetting} and Fig.~\ref{fig:test_layout}. The angular distribution characteristics are reflected in the $\mathbf{pdf}$ shown in Fig.~\ref{fig:test_FP}.
\begin{table} 
    \setlength{\abovecaptionskip}{0cm}
    \setlength{\belowcaptionskip}{-0.2cm}
    \centering
    \caption{\label{tab:QuaDRiGaSetting}Basic parameter settings used in the simulation.}
    \renewcommand\arraystretch{0.9}
    \resizebox{0.49\textwidth}{!}{
    \begin{tabular}{ll}
        \toprule
        \multirow{4}{*}{Antenna setting} & 32 ULA antennas at BS with spacing \\
        & of 0.5 wavelength, centered at (0,0) and \\
        & arranged in a north-south direction (Y-axis) \\
        & Single antenna at UE \\ \midrule

        Center frequency & 2.6 GHz \\
        Bandwidth & 20 MHz \\ \midrule

        \multirow{2}{*}{Scene configuration} & 3GPP-38.901-UMi-LOS \\
        & 3GPP-38.901-UMi-NLOS \\ \midrule

        \multirow{2}{*}{Height} & 10 m for the BS \\
        & 1.5 m for the UE \\ \midrule

        \multirow{2}{*}{Position of sampling area} & Randomly distributed in the 400 m $\times$ 400 m \\
        & area centered relative to the BS \\ \midrule

        \multirow{3}{*}{District range} & Circle with a radius of 5 m for Learnware training set \\
        & Rectangles of various lengths (5 m $\times$ [5,10, \\
        & 20,30,40,50] m) for target test set \\ \midrule

        Sampling number per dataset & 10,000 \\
        Maps number / Learnware number & 144 \\ \midrule

        \multirow{2}{*}{CSI Pretreatment} & SVD \\
        & Normalized to real values in the range $[0,1]$ \\
        \bottomrule
    \end{tabular}}
    \vspace{-5mm}
\end{table}
\begin{table} 
    \centering
    \setlength{\abovecaptionskip}{0cm}
    \setlength{\belowcaptionskip}{-0.2cm}
    \caption{Autoencoder Structures for CNN and Transformer-based CSI Feedback}
    \resizebox{0.49\textwidth}{!}{%
    \renewcommand\arraystretch{0.85}
    \begin{tabular}{lllll}
\toprule
\textbf{Layer} & \textbf{Type} & \textbf{Input} & \textbf{Output} & \textbf{Details} \\
\midrule
\multicolumn{5}{l}{\textit{\textbf{CNN-based Autoencoder~\cite{wen2018deep}}}} \\
\midrule
Input & - & (32, 2) & (32, 2) & Real and imaginary parts \\
\midrule
\multirow{4}{*}{Encoder} 
 & Conv1D & (32, 2) & (32, 2) & Kernel: $7 \times 1$, BN, ReLU \\
 & Reshape & (32, 2) & (64,) & Flatten \\
 & Dense+Sigmoid & (64,) & \textcolor{black}{(4,)} & Compression ratio \textcolor{black}{$CR=1/16$} \\ 
 & Quantization & \textcolor{black}{(4,)} & \textcolor{black}{(16,)} & $Q=4$ bits, \textcolor{black}{feedback\_bits=16} \\
\midrule
\multirow{6}{*}{Decoder} 
& Dequantization & \textcolor{black}{(16,)} & \textcolor{black}{(4,)} & --\\
& Dense & \textcolor{black}{(4,)} & (64,) & -- \\
 & Reshape & (64,) & (32, 2) & -- \\
 & \begin{tabular}[l]{@{}l@{}}RefineNet \\ Block $\times 2$ \end{tabular} & (32, 2) & (32, 2) & \begin{tabular}{@{}l@{}}Conv1D: (32, 2) $\rightarrow$ (32, 8) \\ Conv1D: (32, 8) $\rightarrow$ (32, 16) \\ Conv1D: (32, 16) $\rightarrow$ (32, 2) \\ Skip Connection \end{tabular} \\
 & Conv1D & (32, 2) & (32, 2) & Kernel: $7 \times 1$, BN, ReLU \\
 & Sigmoid & (32, 2) & (32, 2) & Output activation \\
\midrule
\multicolumn{5}{l}{\textit{\textbf{Transformer-based Autoencoder}}~\cite{han2024ai}} \\
\midrule
\multirow{3}{*}{Input} 
 & Input & (32, 2) & (32, 2) & $N_{\mathrm{Token}}=32$, $feature\_dim$=2 \\
 & Linear Embedding & (32, 2) & (32, 81) & $Embedding\_dim=81$, Dense \\
 & Positional Encoding & (32, 81) & (32, 81) & Sinusoidal encoding \\
\midrule
\multirow{5}{*}{Encoder} 
 & \begin{tabular}[l]{@{}l@{}}Transformer Block \\ $\times L_{\mathrm{TF}}$ ($L_{\mathrm{TF}}=12$) \end{tabular} & (32, 81) & (32, 81) & \begin{tabular}{@{}l@{}}Multi-Head Attn: $N_{\mathrm{head}}=3$, Dropout(0.1) \\ LayerNorm, Residual connection \\ FFN: $324 \rightarrow 81$ (4$\times$), ReLU, Dropout(0.1) \\ LayerNorm, Residual connection \end{tabular} \\
 & LayerNorm & (32, 81) & (32, 81) & Final normalization \\
 & Dense & (32, 81) & (32, 2) & Project back to 2 channels \\
 & Flatten & (32, 2) & (64,) & -- \\
 & Dense+Sigmoid & (64,) & \textcolor{black}{(4,)} & \textcolor{black}{$CR=1/16$} \\
 & Quantization & \textcolor{black}{(4,)} & \textcolor{black}{(16,)} & $Q=4$ bits, \textcolor{black}{feedback\_bits=16} \\
\midrule
\multirow{7}{*}{Decoder} 
 & Dequantization & \textcolor{black}{(16,)} & \textcolor{black}{(4,)} & -- \\
 & Dense & \textcolor{black}{(4,)} & (64,) & -- \\
 & Reshape & (64,) & (32, 2) & -- \\
 & Linear Embedding & (32, 2) & (32, 81) & Dense(81), scale by $\sqrt{81}$ \\
 & Positional Encoding & (32, 81) & (32, 81) & Sinusoidal encoding \\
 & \begin{tabular}[l]{@{}l@{}}Transformer Block \\ $\times L_{\mathrm{TF}}$ ($L_{\mathrm{TF}}=12$) \end{tabular} & (32, 81) & (32, 81) & \begin{tabular}{@{}l@{}}Multi-Head Attn: $N_{\mathrm{head}}=3$, Dropout(0.1) \\ LayerNorm, Residual connection \\ FFN: $324 \rightarrow 81$ (4$\times$), ReLU, Dropout(0.1) \\ LayerNorm, Residual connection \end{tabular} \\
 & LayerNorm & (32, 81) & (32, 81) & Final normalization \\
 & Dense & (32, 81) & (32, 2) & Output projection \\
\bottomrule
\end{tabular}}
    \label{tab:network_structure}
    \vspace{-6mm}
\end{table}
\begin{table*}[htbp]
\centering
\caption{Average SGCS performance of scene-specific small models and large general model with various complexity. CR=1/16 and feedback quantization bits $Q$=4. }
\resizebox{0.98\textwidth}{!}{%
\renewcommand\arraystretch{1}
\begin{threeparttable}
\begin{tabular}{c cccccccc cccccccc}
\toprule
\multirow{3}{*}{\begin{tabular}[c]{@{}c@{}}Sample Range \\ ($\mathrm{m}^2$)\end{tabular} }& \multicolumn{8}{c}{LOS} & \multicolumn{8}{c}{NLOS} \\
 \cmidrule(rl){2-9} \cmidrule(rl){10-17} 
 
 & \multicolumn{2}{c}{Scene-specific}
 & \multirow{2}{*}{\begin{tabular}[c]{@{}c@{}}Learnw. Model \\Repository\end{tabular}} & \multicolumn{5}{c}{General Model (various complexity)\tnote{1}} 
 & \multicolumn{2}{c}{Scene-specific}
 & \multirow{2}{*}{
 \begin{tabular}[c]{@{}c@{}}Learnw. Model \\Repository\end{tabular}} & \multicolumn{5}{c}{General Model (various complexity)} \\
 \cmidrule(rl){2-3} \cmidrule(rl){5-9} \cmidrule(rl){10-11}\cmidrule(rl){13-17}
 & \textbf{CNN}\tnote{2} & TF-1 &   & CNN & TF-1 & TF-2 & TF-3 & \textbf{TF-4}\tnote{3} & \textbf{CNN} & TF-1 &  & CNN & TF-1 & TF-2 & TF-3 & \textbf{TF-4} \\
 
 \midrule

$5\times 5$  & 0.915 & 0.915 & 0.889 &0.332  &0.628  &0.725  &0.755  & 0.777       & 0.350 & 0.357 & 0.317 &0.181  &0.208  &0.207  &0.214 & 0.230 \\
$5\times 10$ & 0.881 & 0.883 & 0.855 &0.340  &0.701  &0.719  &0.746  & 0.715       & 0.347 & 0.343 & 0.321 &0.225  &0.213  &0.220  &0.245  & 0.263 \\
$5\times 20$ & 0.879 & 0.886 & 0.808 &0.297  &0.703  &0.710  &0.713  & 0.715       & 0.330 & 0.347 & 0.314 &0.162  &0.205  &0.205  &0.216  & 0.221 \\
$5\times 30$ & 0.868 & 0.881 & 0.721 &0.292  &0.651  &0.684  &0.691  & 0.660       & 0.324 & 0.339 & 0.303 &0.187  &0.213  &0.207  &0.214  & 0.213 \\
$5\times 40$ & 0.848 & 0.858 & 0.703 &0.290  &0.521  &0.579  &0.598  & 0.613       & 0.334 & 0.350 & 0.285 &0.189  &0.217  &0.219  &0.230  & 0.238 \\
$5\times 50$ & 0.844 & 0.865 & 0.652 &0.284  &0.489  &0.491  &0.467  & 0.492       & 0.276 & 0.286 & 0.232 &0.126  &0.189  &0.188  &0.188  & 0.194 \\ \midrule
Parameters [$10^3$] & 2.01 & 58.66 & 2.01 & 2.01 & 58.66 & 424.29 & 1,069.56 & 2,114.99 & 2.01 & 58.66 & 2.01 & 2.01 & 58.66 & 424.29 & 1,069.56 & 2,114.99 \\
Compare with CNN  & 1$\times$ & 29$\times$ & 1$\times$ & 1$\times$ & 29$\times$ & 211$\times$ & 532$\times$ & 1,051$\times$ & 1$\times$ & 29$\times$ & 1$\times$ & 1$\times$ & 29$\times$ & 211$\times$ & 532$\times$ & 1,051$\times$\\ \midrule
FLOPs [$10^6$] & 0.11 & 12.64 & 0.11 & 0.11 & 12.64 & 104.03 & 312.06 & 624.09 & 0.11 & 12.64 & 0.11 & 0.11 & 12.64 & 104.03 & 312.06 & 624.09\\
Compare with CNN  & 1$\times$ & 115$\times$ & 1$\times$ & 1$\times$ & 115$\times$ & 945$\times$ & 2,837$\times$ & 5,673$\times$ & 1$\times$ & 115$\times$ & 1$\times$ & 1$\times$ & 115$\times$ & 945$\times$ & 2,837$\times$ & 5,673$\times$\\
\bottomrule
\end{tabular}
	\begin{tablenotes}
       \footnotesize
       \item[1] Architectures in Table~\ref{tab:network_structure}: CNN (same with Scene-specific and Learnware Model Repository); TF-1: Transformer with $L_{\mathrm{TF}}=2$ and $Embedding\_dim=27$; TF-2: Transformer with $L_{\mathrm{TF}}=2$ and $Embedding\_dim=81$; TF-3: Transformer with $L_{\mathrm{TF}}=6$ and $Embedding\_dim=81$; TF-4: Transformer with $L_{\mathrm{TF}}=12$ and $Embedding\_dim=81$. 
       \item[2] The bold-highlighted CNN is adopted for scene-specific models as the Upper Bound.
       \item[3] The bold-highlighted TF-4 is adopted as the default NN architecture for the General Model in subsequent experiments.
     \end{tablenotes}
    \end{threeparttable}
}
\label{tab:generalization}\vspace{-3mm}
\end{table*}

\begin{table*}[htbp]
\centering
\caption{Average SGCS performance of methods with various CR and feedback quantization bits. Sampling range: $5\times 5\,m^2$.}
\resizebox{0.98\textwidth}{!}{%
\renewcommand\arraystretch{1}
\begin{threeparttable}
\begin{tabular}{cc ccccc ccccc}
\toprule
\multicolumn{2}{c}{Configurations}& \multicolumn{5}{c}{LOS scenes} & \multicolumn{5}{c}{NLOS scenes} \\
 \cmidrule(rl){1-2} \cmidrule(rl){3-7} \cmidrule(rl){8-12} 
 \multirow{2}{*}{CR} & \multirow{2}{*}{\begin{tabular}[c]{@{}c@{}}Quantization\\ bits ($Q$)\end{tabular}} & \multirow{2}{*}{\begin{tabular}[c]{@{}c@{}}General \\Model\end{tabular}} & \multirow{2}{*}{\begin{tabular}[c]{@{}c@{}}Model \\Switch\end{tabular}} & \multirow{2}{*}{\begin{tabular}[c]{@{}c@{}}Learnware Model \\ Repository\end{tabular}}
 & \multirow{2}{*}{\begin{tabular}[c]{@{}c@{}}Original Model \\ Repository\end{tabular}} & \multirow{2}{*}{\begin{tabular}[c]{@{}c@{}}Upper \\Bound\end{tabular}} & \multirow{2}{*}{\begin{tabular}[c]{@{}c@{}}General \\Model\end{tabular}} & \multirow{2}{*}{\begin{tabular}[c]{@{}c@{}}Model \\Switch\end{tabular}} & \multirow{2}{*}{\begin{tabular}[c]{@{}c@{}}Learnware Model \\ Repository\end{tabular}}
 & \multirow{2}{*}{\begin{tabular}[c]{@{}c@{}}Original Model \\ Repository\end{tabular}} & \multirow{2}{*}{\begin{tabular}[c]{@{}c@{}}Upper \\Bound\end{tabular}}\\ \\ \midrule
 \multirow{4}{*}{ 1/16} 
 & 2 & 0.681 & 0.371 & 0.832 & 0.841 & 0.878 & 0.219 & 0.197 & 0.296 & 0.309 & 0.336 \\
 & 3 & 0.715 & 0.453 & 0.850 & 0.859 & 0.881 & 0.230 & 0.222 & 0.313 & 0.328 & 0.356\\
 & 4 & 0.777 & 0.477 & 0.889 & 0.891 & 0.915 & 0.230 & 0.230 & 0.317 & 0.333 & 0.350\\
 & 5 & 0.781 & 0.482 & 0.899 & 0.907 & 0.921 & 0.232 & 0.233 & 0.320 & 0.335 & 0.356 \\ \midrule
1/32 & \multirow{3}{*}{4} & 0.348 & 0.249 & 0.794 & 0.816 & 0.852 & 0.134 & 0.144 & 0.215 & 0.239 & 0.251\\
1/16 & & 0.777 & 0.477 & 0.889 & 0.891 & 0.915 & 0.230 & 0.230 & 0.317 & 0.333 & 0.350\\
1/8  & & 0.843 & 0.681 & 0.897 & 0.906 & 0.920 & 0.420 & 0.385 & 0.472 & 0.489 & 0.510 \\

\bottomrule
\end{tabular}
    \end{threeparttable}
}
\label{tab:quantization}\vspace{-3mm}
\end{table*}
\subsubsection{NN Training and Evaluation}

Each Learnware was trained using a lightweight CNN-based autoencoder (CsiNet~\cite{wen2018deep} with 2D convolutions replaced by 1D, kernel size expanded from 3 to 7). The General Model follows the transformer-based EVCsiNet-T~\cite{han2024ai} (details in Tab.~\ref{tab:network_structure}), aligning with the 3GPP R18 AI/ML study~\cite{han2024ai, 3gpp-rp-213599}.
The dataset was partitioned into training (85\%), validation (10\%), and test (5\%) sets. We used the Adam optimizer with a learning rate of 1e-4, training for 1,000 epochs and saving the model at the point of optimal validation performance.

For targeted testing, QuaDRiGa generated CSI data matching the semantic configurations. Test set sampling areas were rectangles of various lengths (5 m $\times$ [5, 10, 20, 30, 40, 50] m) with random centers and rotations under the 3GPP-38.901-UMi LOS/NLOS configuration. Our solution retrieves the best Learnware based on statistical specification similarity and returns it for optional fine-tuning. We evaluated both the initial and fine-tuned performance.
\begin{figure*}
    \centering
    \setlength{\abovecaptionskip}{-1.5mm}
    \setlength{\belowcaptionskip}{-1cm}
    \subfigure[Testing LOS scene.\label{fig:LOS_test_performance}]{\includegraphics[width=0.4\linewidth]{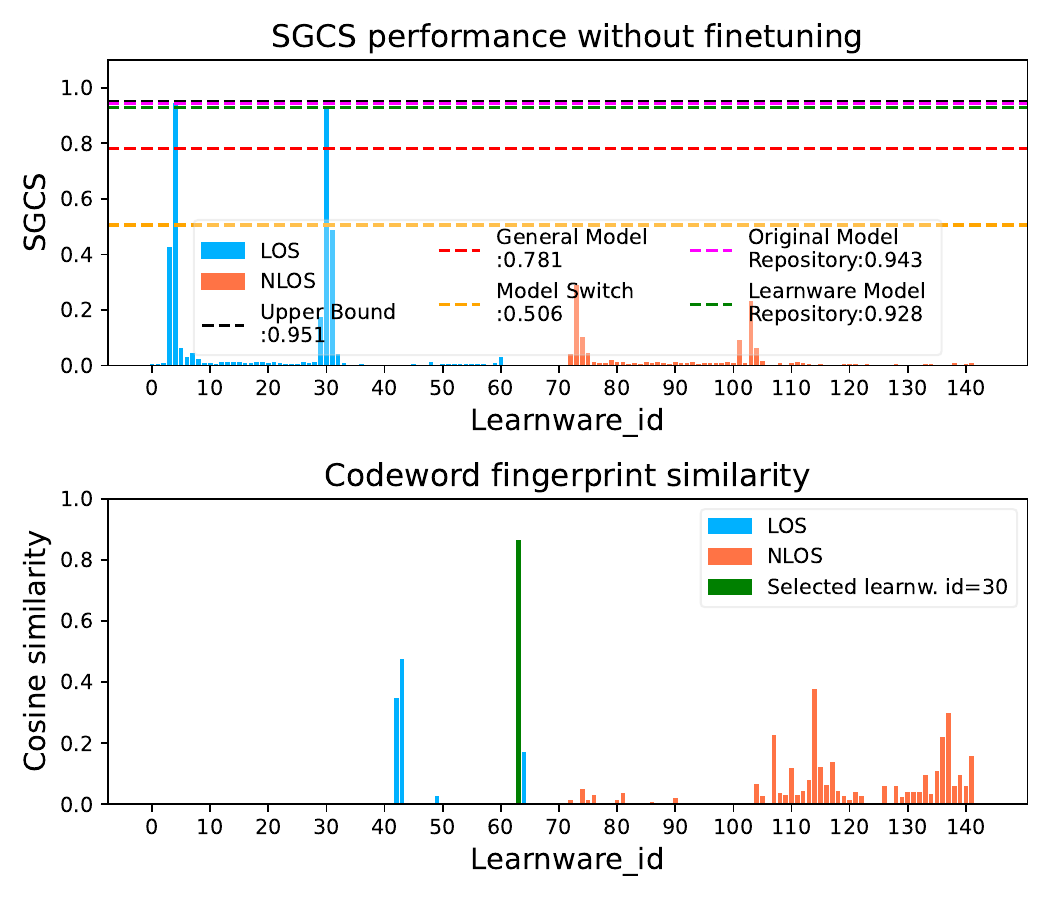}}
    \subfigure[Testing NLOS scene.\label{fig:NLOS_test_performance}]{\includegraphics[width=0.4\linewidth]{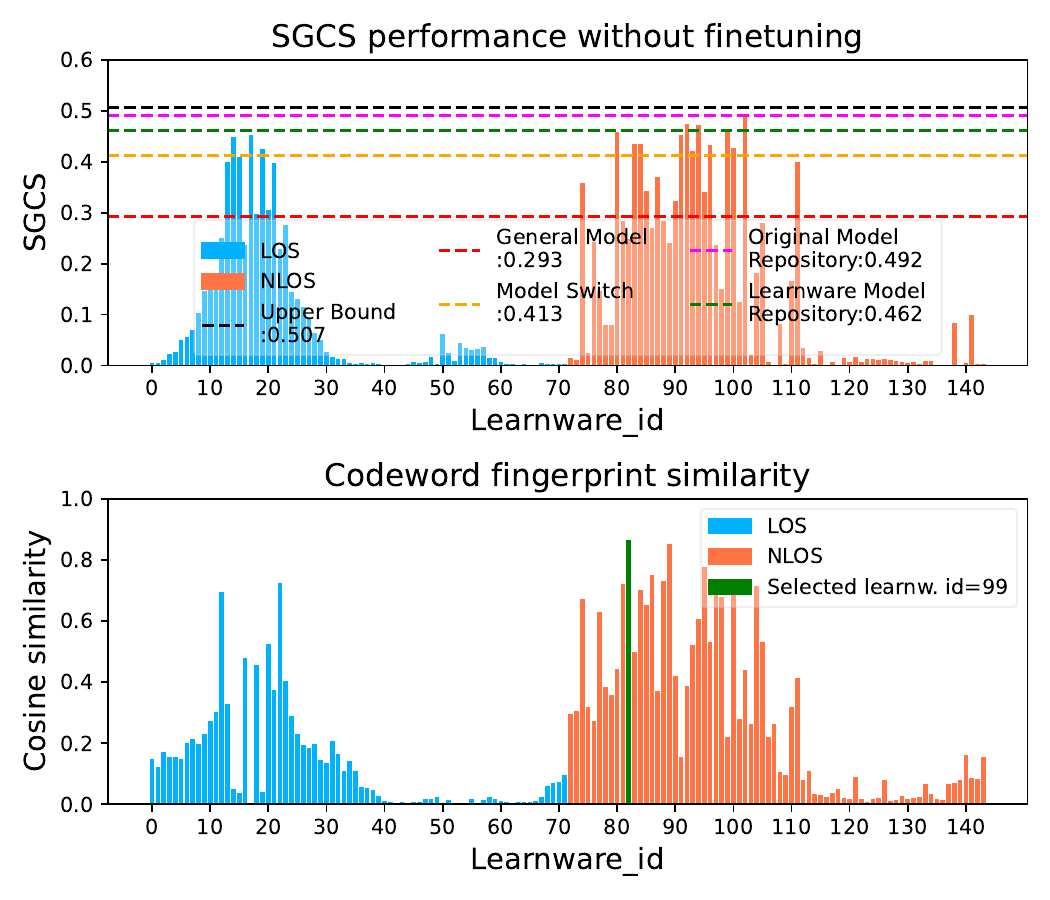}}
    \caption{SGCS performance without fine-tuning and codebook fingerprint cosine similarity. The LOS and NLOS target testing scenes are presented in (a) and (b). In each subfigure, the LOS and NLOS Learnwares are displayed in blue and orange, respectively. The Learnware selected by the codebook fingerprint (i.e., the one with the maximum cosine similarity) is marked in green. The CSI sampling area for the test set is set to $5\,\text{m} \times 5\,\text{m}$, the same as the Learnware training set.  CR=1/16 and feedback quantization bits $Q$=4.}
    \label{fig:Test_performance_similarity}
    \vspace{-5mm}
\end{figure*}

\subsection{Performance Gain}
Performance evaluations are conducted for the aforementioned schemes, all sharing the same architectural design but differing in their training processes:

\begin{itemize}   
    \item \textbf{General Model (with/without fine-tuning):}  
    A transformer-based (architecture in Tab.~\ref{tab:network_structure}) model with high generalization capability is trained using a mixture of the 144 datasets from the training Learnwares. 

    \item \textbf{Model Switch:}  
    Four configuration-specific models are prepared with CNN-based architecture in Tab.~\ref{tab:network_structure}. The 144 datasets are divided into four groups based on LOS/NLOS conditions and the upper/lower sides of the ULA axis at the BS (north/south sides in Fig.~\ref{fig:test_layout}, corresponding to the first/second halves in Fig.~\ref{fig:test_FP}). Each model is trained with a mix of 36 datasets from each group.

    \item \textbf{Original Model Repository:}  
    The performance of the CNN-based 144 models in the repository is evaluated to select the best-performing model, which represents the upper bound of performance for model repository-based schemes.

    \item \textbf{Learnware Model Repository:}  
    The codebook fingerprint sample frequency $\mathbf{pdf}$ is adopted as the statistical specification, with $N$ set to 64. The model library computes the cosine similarity of $\mathbf{pdf}$ between all Learnwares within the specified island\footnote{Given the limited number of Learnwares in this experiment, techniques such as LSH will be implemented to accelerate retrieval when scaling up.} and the target scene test set, selecting the Learnware with the highest similarity.

    \item \textbf{Upper Bound:}  
    With sufficient local data, smaller models often achieve better overfitting performance than generalized models, thus setting the performance upper bound. Due to the large sampling area in the test scenario, datasets ranging from 10,000 to 50,000 samples are used for training to obtain this upper bound.
\end{itemize}

\subsubsection{Initial Performance Verification}

We first evaluate performance without fine-tuning. Tab.~\ref{tab:generalization} first validates the strong performance of the scene-specific model compared to General Model, where the SGCS performance is the mean of the 10 cases included in each sampling area range. 
Specific CNN models (2k parameters, 0.11M FLOPs) are \textcolor{black}{compared against} transformer-based General Models with increasing complexity (TF-1 to TF-4, with \textcolor{black}{millions of} parameters). Results reveal that specific models consistently outperform all General Models across all scenarios, even when General Model capacity expands to 1,000$\times$ the parameter count and 5,600$\times$ the computational cost. This demonstrates that specific advantages persist over an extremely wide capacity range, with million-parameter models still failing to match small specialized models. The diminishing returns in General Model performance also highlight the inefficiency of simply scaling up model capacity. Moreover, deploying such large models imposes significant hardware demands and inference latency, making them impractical for resource-constrained edge devices, whereas lightweight scene-specific models enable efficient on-device deployment.

Tab.~\ref{tab:quantization} analyzes the sensitivity of all methods to different compression ratios and quantization bits. The Learnware Model Repository consistently outperforms the General Model across all configurations, with SGCS gains of 0.05--0.45 (6.4\%--128.2\%) in LOS and 0.05--0.09 (12.4\%--60.4\%) in NLOS. Even under the most constrained feedback (CR=$1/32$, $Q=4$), the Learnware achieves 0.794 (LOS) and 0.215 (NLOS), far exceeding the General Model (0.348 LOS, 0.134 NLOS), demonstrating robustness to feedback payload variations.

Fig.~\ref{fig:Test_performance_similarity} displays representative individual cases randomly selected from the 10 LOS and 10 NLOS test scenarios with $5\mathrm{m}\times5\mathrm{m}$ sampling area. In each subfigure, the blue bars represent the SGCS performance of all 144 LOS Learnware models when tested on that specific LOS/NLOS test scenario, while orange bars represent NLOS Learnware models. 
The green-highlighted bar indicates the Learnware model selected by our codebook fingerprint method (i.e., the model whose training dataset's $\mathbf{pdf}$ has the highest cosine similarity with the test scenario's $\mathbf{pdf}$). The horizontal dashed lines show the performance of five methods evaluated on the test scenario.
The performance ranking consistently follows: Upper Bound $>$ Original Model Repository $>$ Learnware Model Repository $>$ General Model/Model Switch.
In both scenarios, repository-based selections closely approach the Upper Bound, significantly outperforming General Model \textcolor{black}{by approximately 18.8\% ($0.147/0.781$) in LOS and 57.7\% ($0.169/0.293$) in NLOS}, demonstrating the effectiveness of \textcolor{black}{$\mathbf{pdf}$-based} model selection across different propagation conditions.
Extensive testing reveals that Learnware selection is robust to LOS/NLOS conditions: green bars (selected models) predominantly appear among blue bars (LOS Learnwares) for LOS test scenarios, and among orange bars (NLOS Learnwares) for NLOS scenarios. Hence, LOS/NLOS is excluded from semantic specifications.

\textcolor{black}{As shown in Fig.~\ref{fig:Test_performance_similarity}, it is hard to decide from a predefined threshold whether a Learnware requires fine-tuning, because the Upper Bound is relatively high for LOS test scenarios but relatively low for NLOS scenarios, making fine-tuning necessary in the latter.} To improve Learnware performance in NLOS scenarios, one approach is to increase the feedback overhead by switching to a different specification island.

\begin{figure*}
    \centering
    \setlength{\belowcaptionskip}{-1cm}
    \setlength{\abovecaptionskip}{0cm}
    \subfigbottomskip=0mm
    \subfigcapskip=0mm
    \subfigure[Testing LOS scene.\label{fig:LOS_test_finetune}]{\includegraphics[width=0.325\linewidth]{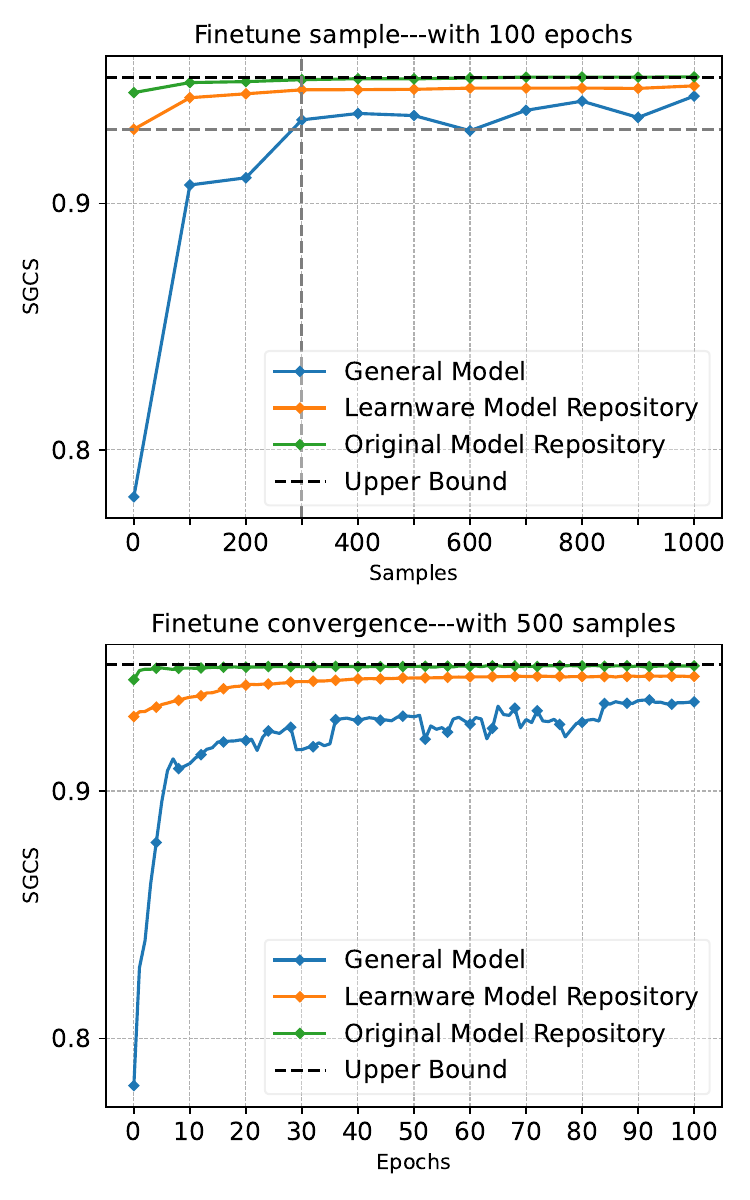}}
    \subfigure[Testing NLOS scene.\label{fig:NLOS_test_finetune}]{\includegraphics[width=0.325\linewidth]{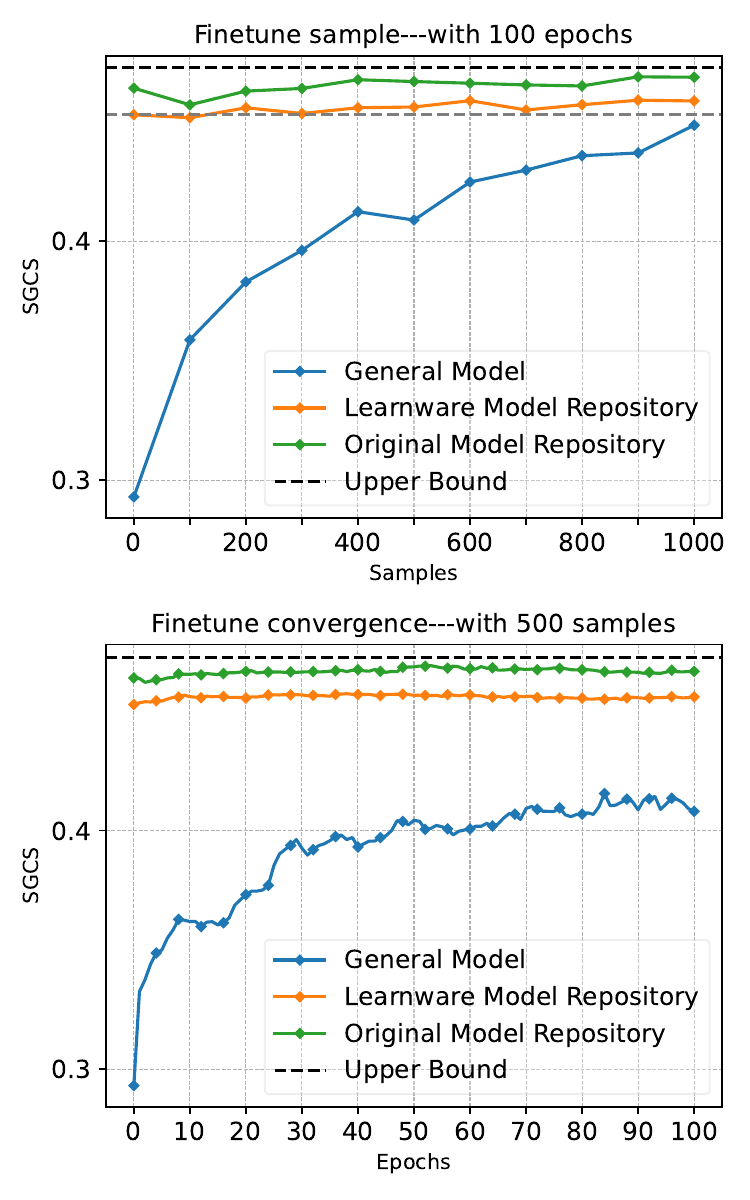}}
    \subfigure[Average sgcs performance.\label{fig:Avr_test_finetune}]{\includegraphics[width=0.325\linewidth]{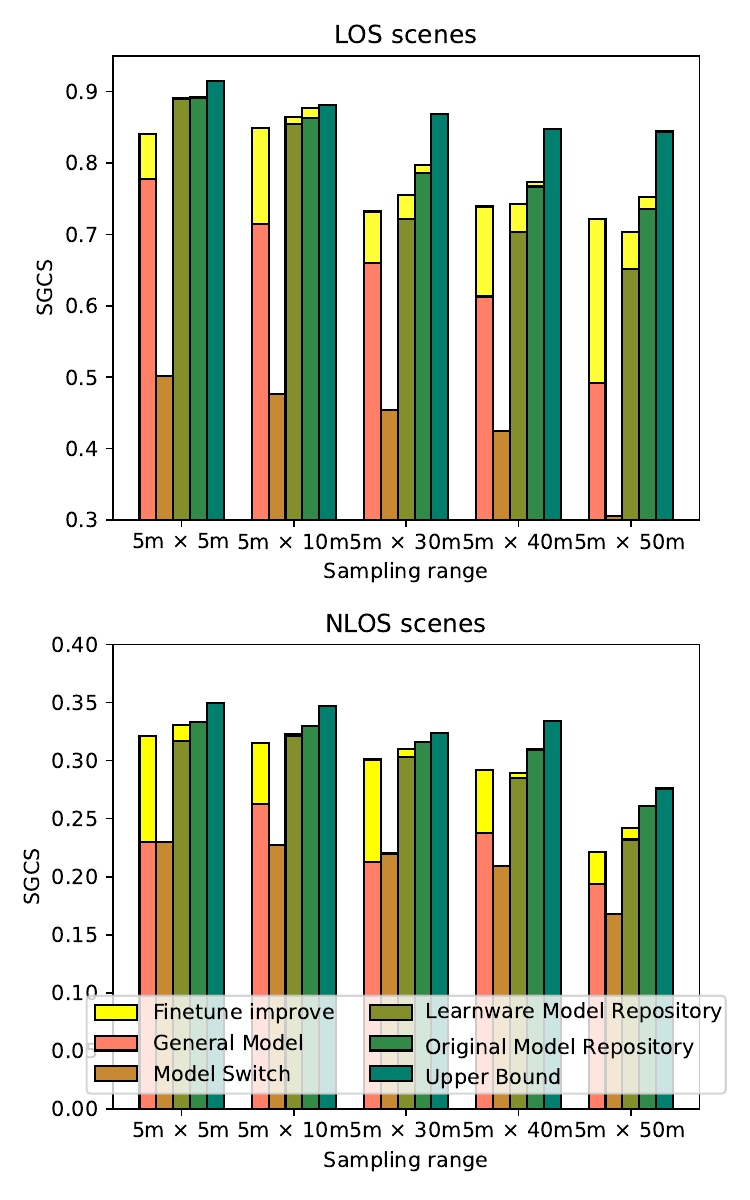}}
    \vspace{2mm}
    \caption{Performance improvements on the two cases from Fig. \ref{fig:Test_performance_similarity} post-fine-tuning. The LOS and NLOS target test scenarios are shown in (a) and (b), respectively. In each subfigure, the upper and lower sections represent the number of samples used for fine-tuning and the number of fine-tuning epochs, respectively. Tested CSI sampling area: $5\,\mathrm{m} \times 5\,\mathrm{m}$. The corresponding fine-tuning results are summarized in Tab.~\ref{tab:Finetune_40m}. (c) presents the average SGCS performance over multiple sampling areas, where each bar represents the average across 10 test scenarios. Fine-tuning overhead: 500 samples and 100 epochs.
    } 
    \label{fig:Test_finetune}
    \vspace{-0.25cm}
\end{figure*}

\begin{table} 
\centering
\caption{SGCS performance before and after fine-tuning (500 samples, 100 epochs) of individual representative case, corresponding to Fig.~\ref{fig:Test_finetune}.}
\label{tab:Finetune_40m}
\resizebox{0.49\textwidth}{!}{%
\renewcommand\arraystretch{1}
\begin{tabular}{ccccc} \toprule
\multirow{4}{*}{Methods} & \multicolumn{2}{c}{LOS Test Scene} & \multicolumn{2}{c}{NLOS Test Scene} \\
\cmidrule(rl){2-3} \cmidrule(rl){4-5} 
& \begin{tabular}[c]{@{}c@{}}Without \\ fine-tuning\end{tabular} 
& \begin{tabular}[c]{@{}c@{}}With \\ fine-tuning\end{tabular} 
& \begin{tabular}[c]{@{}c@{}}Without \\ fine-tuning\end{tabular} 
& \begin{tabular}[c]{@{}c@{}}With \\ fine-tuning\end{tabular} \\ \midrule
General Model & 0.781 & 0.936 & 0.293 & 0.408 \\
Model Switch         & 0.506 & None & 0.413 & None \\
Original Model Repository  & 0.943 & 0.951 & 0.492 & 0.499 \\
Learnware Model Repository & 0.928 & 0.946 & 0.462 & 0.472 \\
Upper Bound & \multicolumn{2}{c}{0.951} & \multicolumn{2}{c}{0.507} \\ \bottomrule
\end{tabular}%
} \vspace{-1mm}
\end{table}

\subsubsection{Fine-Tuning Performance}
The convergence and sample requirements for fine-tuning are depicted in Figs.~\ref{fig:LOS_test_finetune} and \ref{fig:NLOS_test_finetune} (same cases from Fig.~\ref{fig:Test_performance_similarity}), with detailed SGCS performance before and after fine-tuning provided in Tab.~\ref{tab:Finetune_40m}. 
The results demonstrate that minimal fine-tuning of the selected Learnware can effectively boost performance, achieving parity with the Original model repository and, in some cases, even match the Upper bound.
Learnware fine-tuning overhead is minimal, as it provides a better initialization point than General Model. 
For example, in the LOS Finetune-sample scenario, General Model requires 300 samples to match the initial performance of the Learnware after 100 fine-tuning epochs (indicated by the gray dashed lines), meaning that direct Learnware deployment saves at least 300 samples times 100 epochs of fine-tuning effort compared to General Model. In the NLOS scenario, the General Model needs up to 1000 samples over 100 epochs to catch up with the Learnware's initial performance. 
This demonstrates that direct Learnware deployment saves \textcolor{black}{300 to 1000} training samples \textcolor{black}{(depending on propagation conditions)} compared to fine-tuning the General Model from scratch, while achieving superior final performance after fine-tuning.

We also conducted experiments on multiple test scenarios across different sampling areas, each consisting of 10 LOS and 10 NLOS cases. The averaged performance is presented in Fig.~\ref{fig:Avr_test_finetune}. The fine-tuning overhead was set to 500 samples for 100 epochs, with yellow bars indicating the corresponding fine-tuning gain. The results confirm that repository-based approaches provide superior initialization, and in some cases, outperform General Model even after its fine-tuning.
Additionally, the slower performance improvement observed for Learnware compared to General Model is due to its weaker generalization ability and its closer proximity to the upper bound. This proximity makes it more challenging to reach the optimal solution and slows down the gradient descent process.
 
\begin{table*}[t]
\setlength{\abovecaptionskip}{0cm}
\renewcommand\arraystretch{1}
\centering
\caption{Performance of Multi-Level LSH with Adjustable Search Depth}
\label{tab:lsh_performance}
\resizebox{0.6\textwidth}{!}{
\begin{tabular}{lccccccccc}
\toprule
\textbf{Level} & \textbf{k} & \textbf{L} & \textbf{Input Candidate} & \textbf{LSH Score} & \textbf{Brute-force} & \textbf{Relative} & \textbf{Per-Level} & \textbf{Cumulative} & \textbf{Speedup} \\
 & & & \textbf{Repository Size} & & \textbf{Upper Bound} & \textbf{Accuracy} & \textbf{Time (ms)} & \textbf{Time (ms)} & \textbf{Factor} \\
\midrule
L1 & 4 & 25 & $1.0\times 10^8$ & 0.85 & 0.99 & 85.9\% & 0.25 & 0.25 & 20,000$\times$ \\
L2 & 5 & 18 & $6.25\times 10^6$ & 0.91 & 0.99 & 91.9\% & 0.35 & 0.60 & 8,333$\times$ \\
L3 & 6 & 12 & $1.95\times 10^5$ & 0.96 & 0.99 & 97.0\% & 0.30 & 0.90 & 5,556$\times$ \\
L4 & 7 & 6 & $3.05\times 10^3$ & 0.98 & 0.99 & 99.0\% & 0.20 & 1.10 & 4,545$\times$ \\
Final & - & - & 48 & 0.99 & 0.99 & 100\% & 0.05 & 1.15 & 4,348$\times$ \\
\bottomrule
\end{tabular}
}
\end{table*} 

\begin{table*}[ht!]
    \setlength{\abovecaptionskip}{0cm}
    \renewcommand\arraystretch{1.1}
    \centering
    \caption{\label{tab:complexity}Online deployment cost comparison for the general model, model switch, and proposed model repository-based schemes.}

	\resizebox{0.98\textwidth}{!}{
 	\begin{threeparttable}
    \begin{tabular}{c cccc cccc} \toprule
\multirow{2}{*}{\begin{tabular}[c]{@{}c@{}}Deployment \\ considerations\end{tabular}} 
& \multicolumn{2}{c}{Communication overhead} 
& \multicolumn{2}{c}{Time complexity}    
& \multicolumn{2}{c}{Communication overhead [KB]\tnote{1}} 
& \multicolumn{2}{c}{Time complexity [s]\tnote{2}} \\ 
\cmidrule(rl){2-3} \cmidrule(rl){4-5} \cmidrule(rl){6-7} \cmidrule(rl){8-9}
& Data center-BS  & BS-UE  
& Fine-tuning  & Model retrieval  
& Data center-BS    & BS-UE   
& Fine-tuning & Model retrieval   \\ \midrule

General Model   
& None  & None  
& None  & None  
& 0    & 0     
& 0    & 0     \\

\begin{tabular}[c]{@{}c@{}}General Model \\ + fine-tuning\end{tabular}  
& None  
& \begin{tabular}[c]{@{}c@{}}500 samples for fine-tuning;\\ return 1 encoder\end{tabular}   
& \begin{tabular}[c]{@{}c@{}}100 epochs,\\ 500 samples\end{tabular} 
& None    
& 0    
& 240 + 3.37$\times 10^4$           
& 373.5   
& 0   \\

Model Switch   
& None 
& \begin{tabular}[c]{@{}c@{}}500 samples for evaluation;\\ return 1 encoder\end{tabular} 
& None   
& Evaluate 4 models    
& 0  
& 240 + 5.8   
& 0    
& 0.71  \\

\begin{tabular}[c]{@{}c@{}}Original \\ model repository\end{tabular}   
& \begin{tabular}[c]{@{}c@{}}500 samples for evaluation;\\ return 1 model\end{tabular} 
& \begin{tabular}[c]{@{}c@{}}500 samples for fine-tuning;\\ return 1 encoder\end{tabular}   
& \begin{tabular}[c]{@{}c@{}}100 epochs,\\ 500 samples\end{tabular} 
& Evaluate 144 models  
& 240 + 31.7   
& 240 + 5.8     
& 61.6   
& 25.7  \\

\begin{tabular}[c]{@{}c@{}}Learnware \\ model repository\end{tabular}   
& \begin{tabular}[c]{@{}c@{}}Specification;\\ return 1 model\end{tabular}       
& \begin{tabular}[c]{@{}c@{}}500 samples for fine-tuning;\\ return 1 encoder\end{tabular}   
& \begin{tabular}[c]{@{}c@{}}100 epochs,\\ 500 samples\end{tabular} 
& Retrieve from 144 models 
& 2.6 + 31.7   
& 240 + 5.8   
& 61.6    
& 1.48e-5  \\ \bottomrule              
\end{tabular}

	\begin{tablenotes}
       \footnotesize
       \item[1] File sizes: 500 samples = 240 KB; CNN autoencoder = 31.7 KB; CNN encoder = 5.83 KB; TF-4 autoencoder = 67.5 MB; TF-4 encoder = 33.7 MB; semantic specification = 0.25 KB; statistical specification = 2.36 KB.
       \item[2] All measurements are obtained using a Tesla V100 GPU. Fine-tuning 500 samples for 100 epochs averaged 373.5 seconds for TF-4 Transformer and 61.6 for CNN. Model evaluation took 0.178 seconds per model. Cosine similarity computation for all 144 Learnwares averaged 1.48e-5 seconds.
     \end{tablenotes}
    \end{threeparttable}}
    \vspace{-5mm}
\end{table*}

\subsection{Adjustable-Speed LSH}
\label{subsection:lsh-experiment}
We evaluate the proposed adjustable-speed multi-level LSH retrieval strategy, validating the accuracy-latency trade-offs and demonstrate the practical feasibility for real-time CSI feedback deployment.
We create a large-scale specification space containing 100 million statistical specification vectors. First, we generate 1,000 CSI datasets using QuaDRiGa simulations under diverse urban microcell scenarios and compute their $\mathbf{pdf}$. We then fit a Gaussian Mixture Model (GMM) with 50 components to these real $\mathbf{pdf}$ vectors and perform data augmentation by sampling 100 million normalized 64-dimensional specification vectors. The repository is organized into four LSH levels, with various $L$ hash tables and $k$-bit signatures, yielding next-level's candidate set sizes of 6.25M, 195k, 3k, and 48 models. We test 1,000 query specifications (from the same GMM distribution) and measure average cosine similarity between retrieved anchors and queries (i.e., LSH score). The brute-force upper bound is computed similarly by exhaustive search over the entire repository (100 million vectors). Relative Accuracy is the ratio of LSH Score to Brute-force Upper Bound. Speedup Factor compares the cumulative LSH retrieval time against brute-force search (5,000 ms for 100 million vectors).

Tab.~\ref{tab:lsh_performance} summarizes the performance of our multi-level LSH strategy across different search depths. Deeper search levels yield higher accuracy but incur greater cumulative latency, enabling users to select the appropriate trade-off based on their requirements. 
Shallower levels (L1-L2) provide ultra-fast retrieval (0.25--0.60 ms) with good accuracy (85.9--91.9\%), while deeper levels (L3-L4) deliver near-optimal accuracy (97.0--99.0\%) at moderately higher latencies (0.90--1.15 ms).
All cumulative latencies (0.25--1.15 ms) remain well within the typical CSI feedback cycles of 10--80 ms in 5G NR systems~\cite{3GPP38214}, confirming the practical feasibility of our approach.

\begin{figure*}[t]
    \centering
    \setlength{\abovecaptionskip}{2mm}
    \subfigbottomskip=0mm
    \subfigcapskip=-3mm
    \subfigure[Performance ranking. \label{fig:ratio_2000tasks}]{\includegraphics[width=0.325\linewidth]{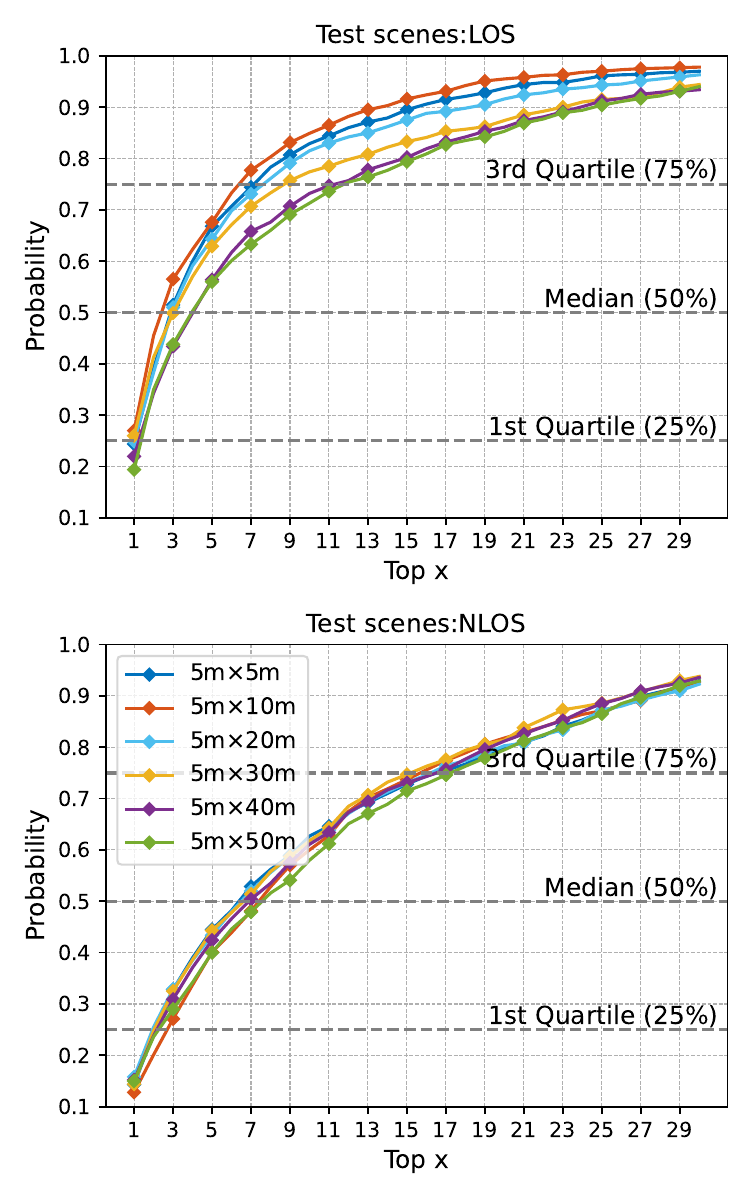}}
    \subfigure[Performance gap boxplot.\label{fig:Sgcs_gap_2000tasks}]{\includegraphics[width=0.325\linewidth]{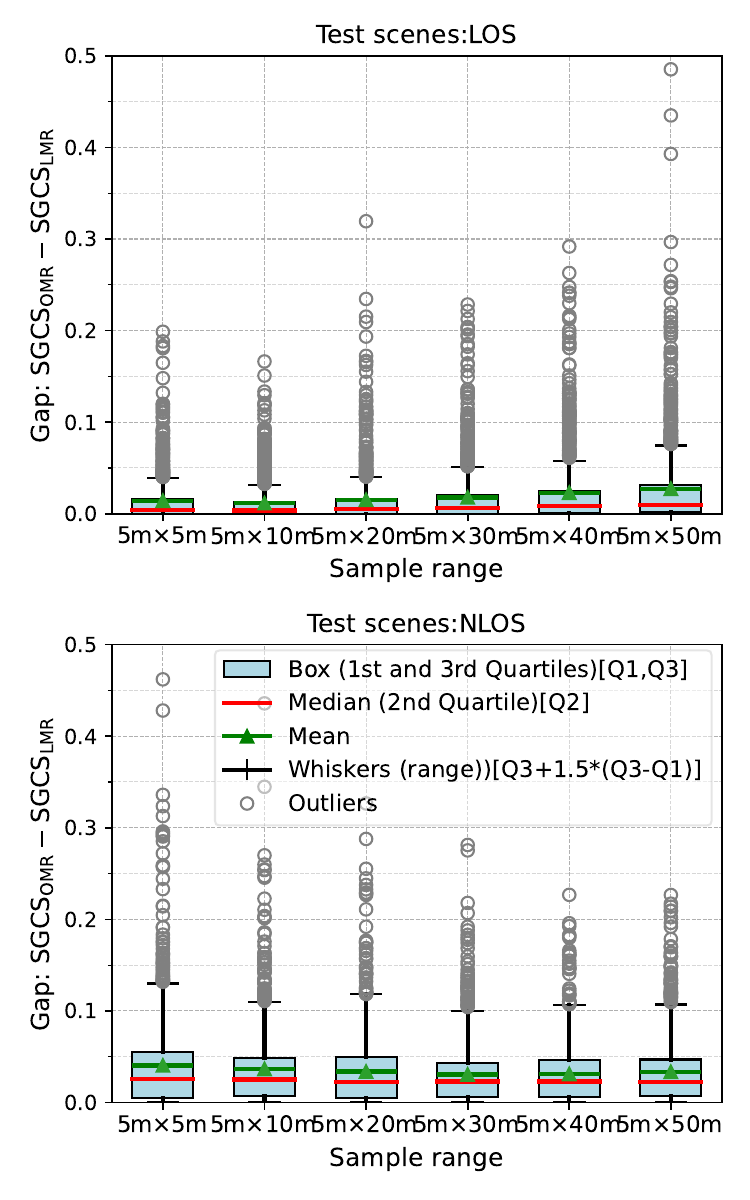}}
    \subfigure[Probability of meeting performance threshold\label{fig:Sgcs_threshold_2000tasks}]{\includegraphics[width=0.325\linewidth]{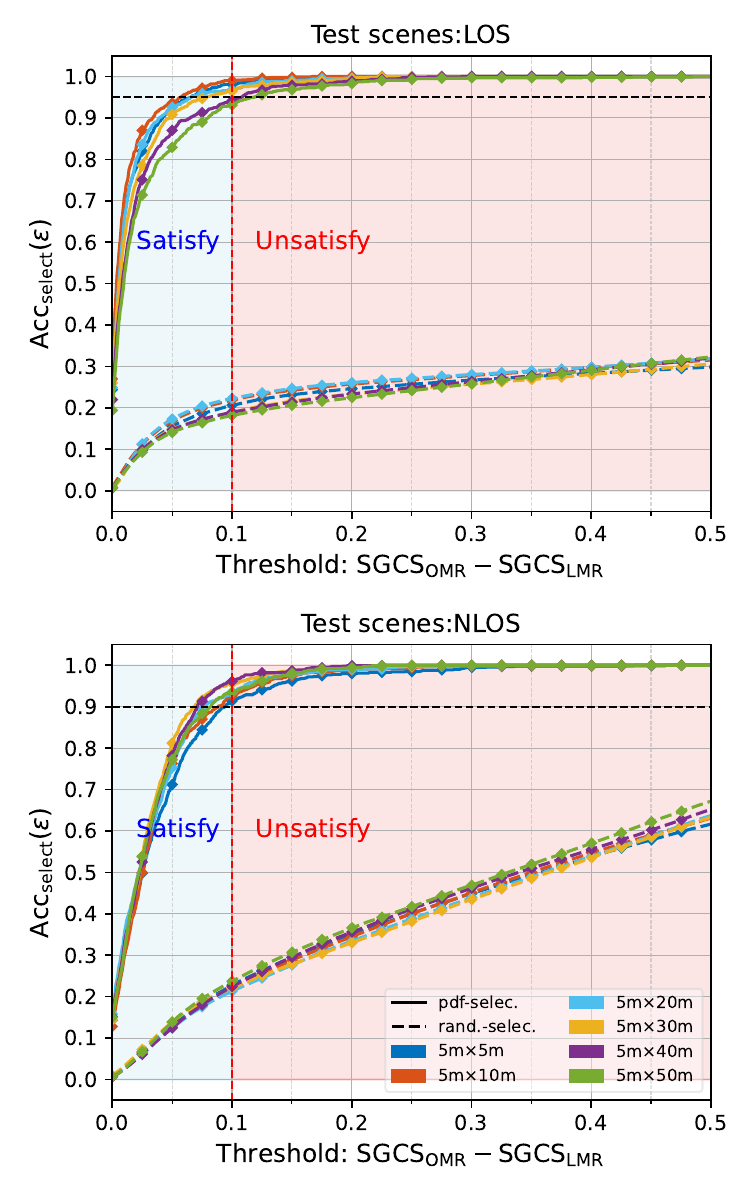}}

    \caption{We simulated 1,000 LOS and 1,000 NLOS scenarios in each sampling area, with each scenario comprising 500 samples. (a) Performance ranking of the learnware selected by $\mathbf{pdf}$ within the model repository (comprising 144 models), i.e., the probability of being ranked in the top x positions; (b) box-plot of the performance gap between the $\mathbf{pdf}$-selected learnware and the top-performing model (i.e., Original model repository); (c) Selection accuracy: probability that the performance gap of the $\mathbf{pdf}$-selected learnware meets the expected threshold.}
    \label{fig:pdf-selection-varify}
    \vspace{-5mm}
\end{figure*}
\subsection{Complexity Analysis}

In Tab.~\ref{tab:complexity}, we quantitatively compare the online deployment costs of General Model, Model Switch, and the proposed model repository-based schemes, including communication overhead and time complexity. Given a fine-tuning budget of 500 samples over 100 epochs, and with model performance evaluation also requiring 500 samples, communication overhead is categorized into two segments: from the AI data center to the BS, and from the BS to the UE. Only the model repository-based schemes incur communication costs in the former. 

The Learnware system addresses the Original model repository's drawback of transmitting local data by significantly reducing transmission overhead from 240 KB to 2.6 KB. Notably, the TF-4 General Model exhibits substantially higher communication costs: its encoder size reaches 33.7 MB, three orders of magnitude larger than the CNN encoder (5.83 KB), making frequent model updates impractical for bandwidth-constrained scenarios.

\begin{figure*}[t]
    \centering
    \setlength{\abovecaptionskip}{2mm}
    \subfigbottomskip=0mm
    \subfigcapskip=-3mm
    \subfigure[Learnware model repository performance with various numbers of learnwares. \label{fig:learnware_num}]{\includegraphics[width=0.315\linewidth]{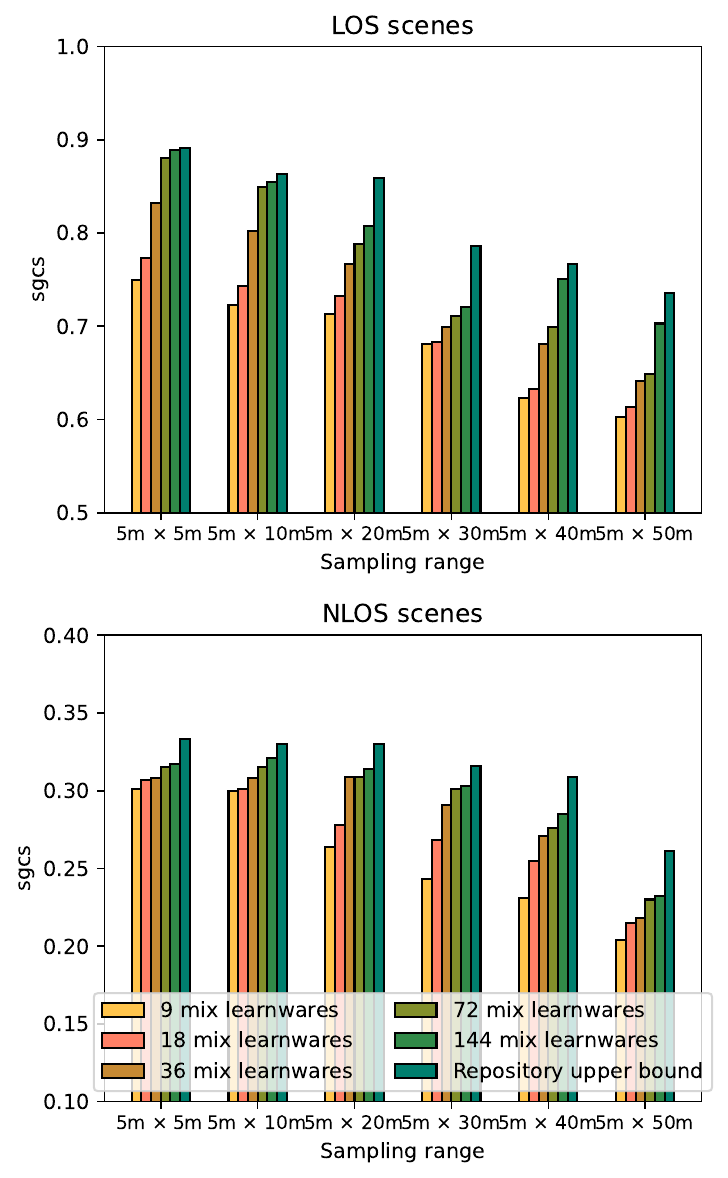}}
    \subfigure[Sample size impacts for $\mathbf{pdf}$ and SGCS in Learnware Model Repository.\label{fig:samplesize_impact}]{\includegraphics[width=0.347\linewidth]{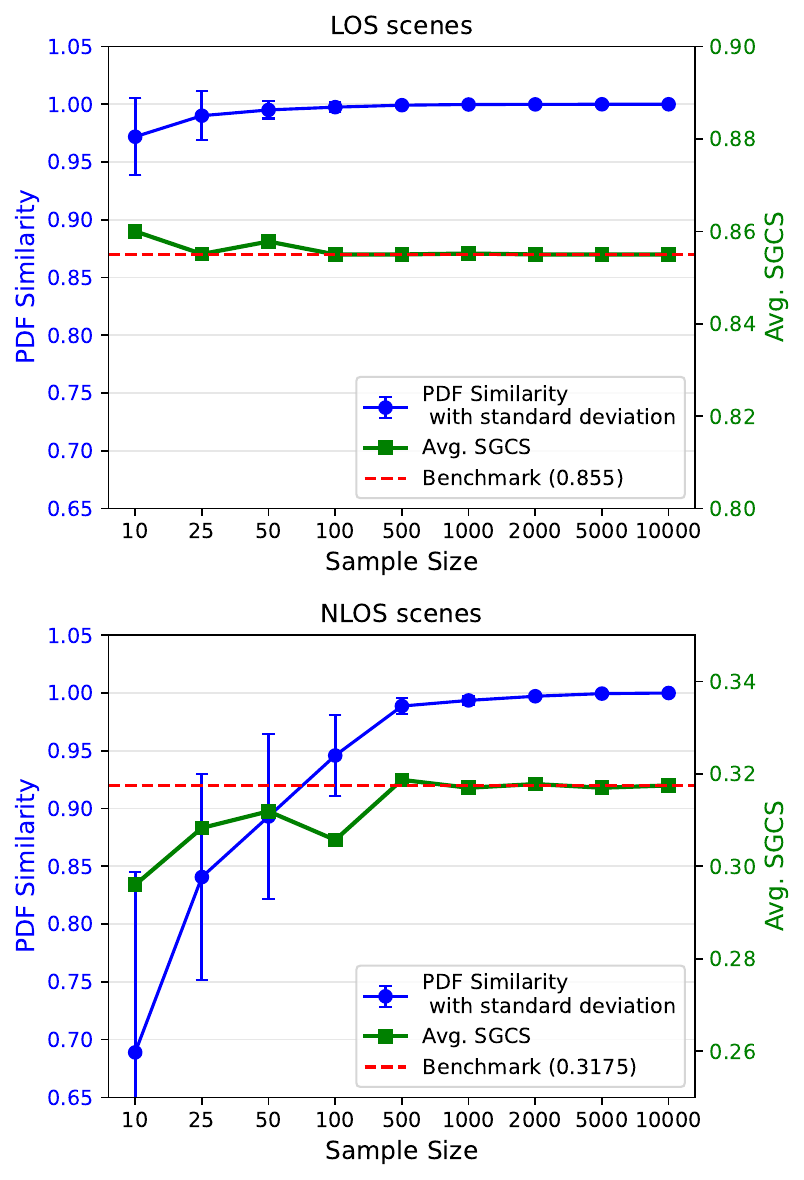}}
    \subfigure[\textcolor{black}{SGCS performance without fine-tuning in OFDM scenario.}\label{fig:multicarrier_perf}]{\includegraphics[width=0.32\linewidth]{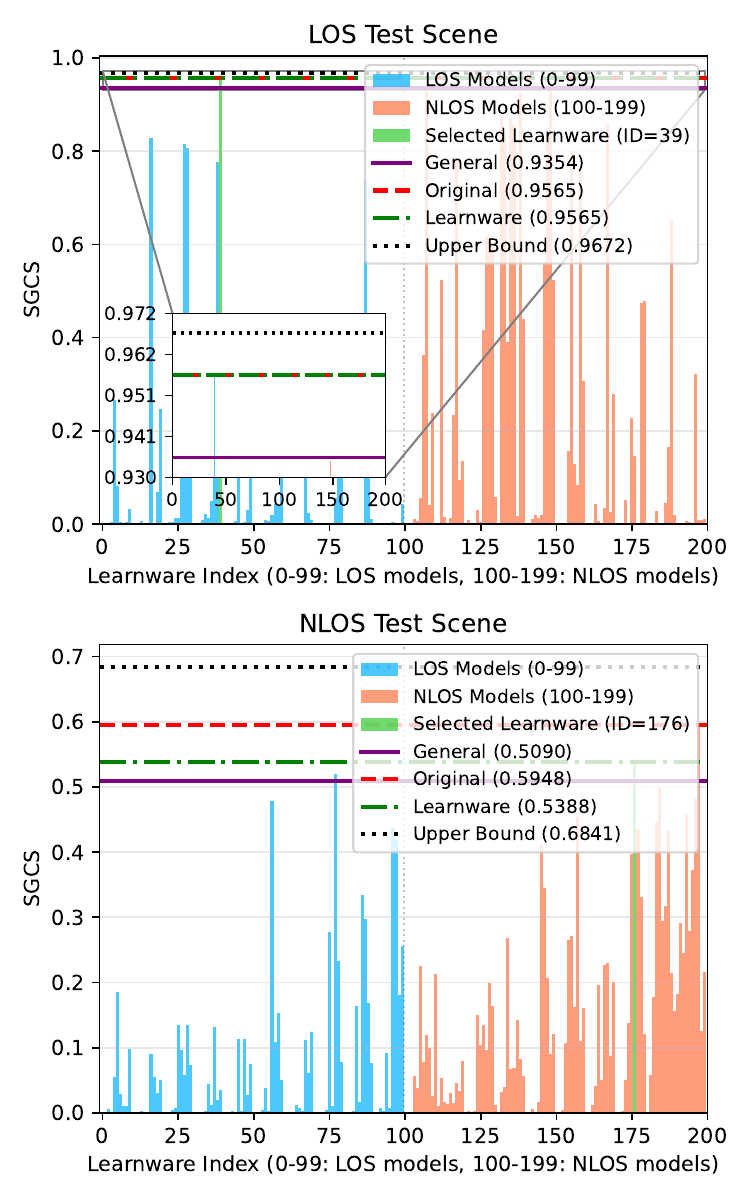}}

    \caption{
     (a) Average SGCS performance of Learnware Model repository, with each group of 10 cases. CR=1/16 and feedback quantization bits $Q$=4; 
    (b) Test scenarios: 10 cases for LOS/NLOS, $5\,m\times 5\,m$ sampling range, For each sample size, the estimated $\mathbf{pdf}$ is compared against the benchmark (full 10,000 samples), Performance loss is defined as SGCS$_{\text{benchmark}} -$ SGCS$_{\text{other}}$. PDF similarity measures the cosine similarity between the estimated $\mathbf{pdf}$ and the benchmark $\mathbf{pdf}$. \textcolor{black}{(c) Multi-carrier OFDM scenario without fine-tuning. The configuration and legend follow the same format as Fig.~\ref{fig:Test_performance_similarity}.}}
    \label{fig:samplesize-num}
    \vspace{-5mm}
\end{figure*}

Time complexity is critical for online deployment and is divided into fine-tuning time and model retrieval time. Fine-tuning the TF-4 Transformer requires 373.5 seconds, over six times longer than the CNN-based model (61.6 seconds), due to its massive parameter count (2.1 million). This becomes a significant disadvantage for the General Model + fine-tuning scheme, as the system must fall back to traditional feedback during this extended period, potentially degrading feedback quality. In contrast, the Learnware model repository requires only 1.48e-5 seconds for retrieval, which corresponds to the cost of computing inner products and is thus negligible. Even when scaled to billion-model repositories, our adjustable-speed multi-level LSH (Algorithm~\ref{alg:lsh}) maintains retrieval latency within 0.25--1.15 ms (Tab.~\ref{tab:lsh_performance}), well below the 10--80 ms CSI feedback cycles specified in 3GPP standards~\cite{3GPP38214}. The near-instantaneous response ensures that during fine-tuning, the Learnware model can be deployed immediately while the locally fine-tuned model trains in the background. This scalability, combined with the minimal communication overhead of statistical specifications, positions the Learnware framework as a practical solution for real-world deployments where both bandwidth and latency are constrained.

\subsection{Further Discussion}

\subsubsection{Validation for $\mathbf{pdf}$ Selection}

To validate the accuracy of codebook fingerprint $\mathbf{pdf}$ selection, we analyzed both the ranking of the Learnware selected by $\mathbf{pdf}$ and its performance gap compared to the top-performing model, as illustrated in Fig.~\ref{fig:pdf-selection-varify}. We simulated 1,000 LOS and 1,000 NLOS test scenarios for each sampling area, with each scenario containing 500 samples (used to generate $\mathbf{pdf}$ and to evaluate performance). This extensive simulation ensures robust probability estimations.

Fig.~\ref{fig:ratio_2000tasks} shows the probability that the test performance of the $\mathbf{pdf}$-selected Learnware ranks among the top $x$ positions in the model repository, which consists of 144 models. In over 90\% of cases, the ranking falls within positions 25 to 27. Additionally, the probability of ranking in the top three is 45\% to 55\% for LOS scenarios and 25\% to 35\% for NLOS scenarios.

A more intuitive indicator is the SGCS performance gap between the $\mathbf{pdf}$-selected Learnware and the top-performing model ($\mathrm{SGCS}_{\mathrm{OMR}} - \mathrm{SGCS}_{\mathrm{LMR}}$). Fig.~\ref{fig:Sgcs_gap_2000tasks} presents a boxplot of the performance gap distribution. The lower boundary, median line, and upper boundary of each box represent the first quartile (Q1), median (Q2), and third quartile (Q3), respectively. The box length reflects the interquartile range of the data. In LOS scenarios, the shorter boxes indicate a more concentrated performance gap distribution, which becomes more dispersed as the sampling area increases. Conversely, in NLOS scenarios, the distribution becomes more concentrated with larger sampling areas, since the performance degradation of the top-performing model offsets the impact of less accurate $\mathbf{pdf}$ selection. The median line is consistently biased towards the lower end of the box, indicating that the performance gap is generally small. The whiskers represent the range of the data and help identify potential outliers. The results show that even in outlier cases, the performance gap does not exceed 0.5.

To quantitatively assess selection accuracy, we define it as the probability that the performance gap falls within a threshold $\epsilon$:
\begin{equation}
\mathrm{Acc}_{\mathrm{select}}(\epsilon) = \Pr\left(|\mathrm{SGCS}_{\mathrm{OMR}} - \mathrm{SGCS}_{\mathrm{LMR}}| \leq \epsilon\right),
\end{equation}
Fig.~\ref{fig:Sgcs_threshold_2000tasks} presents selection accuracy with $\epsilon = 0.1$. The light blue area indicates acceptable cases (gap $\leq 0.1$), while the light red area indicates failure (gap $> 0.1$). The results demonstrate approximately 95\% accuracy for LOS and 90\% for NLOS scenarios, significantly exceeding random selection.

\subsubsection{Impact of Learnware Quantity}

We further investigate the impact of the Learnware model library size on performance. As shown in Fig.~\ref{fig:learnware_num}, we randomly selected varying numbers of Learnwares from a pool of 144 to form model libraries of different sizes. This process was repeated 100 times to reduce the effects of randomness. The mean performance across 10 test scenarios was then computed for each sampling region. The last column in each group of bars represents the performance of the best Learnware among the 144, corresponding to the Original model repository.

The experimental results indicate that as the model library size increases, the performance of the selected Learnwares also improves. However, this improvement exhibits diminishing marginal returns, suggesting that the model library size should be kept within a reasonable range to balance performance with retrieval efficiency. Additionally, as the sampling region of the test scenarios expands, the performance gap between the Original model repository and the selected Learnwares increases, particularly in NLOS scenarios. This is attributed to the more dispersed angular distribution of the datasets, which reduces the effectiveness of $\mathbf{pdf}$ as a statistical specification criterion. Therefore, when deploying the system, the BS should carefully define local scene boundaries to avoid excessively dispersed angular distributions.
\begin{table*}[t]
\caption{Similarity and Distance Metrics for Probability Distributions}
\label{tab:metrics-formular}
\centering
\resizebox{0.8\textwidth}{!}{%
\renewcommand\arraystretch{0.6}
\begin{threeparttable}
\begin{tabular}{@{}lllll@{}}
\toprule
\textbf{Metric} & \textbf{Original Formula} & \textbf{Original Range} & \textbf{Normalized Similarity} & \textbf{Complexity}\tnote{1} \\ \midrule
Cosine Similarity & $\displaystyle s_{\cos}(p,q) = \frac{\sum_{i} p_i q_i}{\|p\| \cdot \|q\|}$ & $[0, 1]$ & $S_{\cos} = s_{\cos}$ & Low \\ \addlinespace
Bhattacharyya Coefficient~\cite{bhattacharyya1943measure} & $\displaystyle s_{\text{Bhatt}}(p,q) = \sum_{i} \sqrt{p_i q_i}$ & $[0, 1]$ & $S_{\text{Bhatt}} = s_{\text{Bhatt}}$ & Medium \\ \addlinespace
Hellinger Distance~\cite{hellinger1909neue} & $\displaystyle d_{\text{Hell}}(p,q) = \frac{1}{\sqrt{2}}\left\|\sqrt{p}-\sqrt{q}\right\|_2$ & $[0, 1]$ & $S_{\text{Hell}} = 1 - d_{\text{Hell}}$ & Medium \\ \addlinespace
Total Variation & $\displaystyle d_{\text{TV}}(p,q) = \frac{1}{2}\sum_{i} |p_i - q_i|$ & $[0, 1]$ & $S_{\text{TV}} = 1 - d_{\text{TV}}$ & Low \\ \addlinespace
Jensen-Shannon Divergence~\cite{lin1991divergence} & $\displaystyle d_{\text{JS}}(p,q) = \frac{1}{2}D_{\text{KL}}(p\|m) + \frac{1}{2}D_{\text{KL}}(q\|m)$ & $[0, 1]$ & $S_{\text{JS}} = 1 - d_{\text{JS}}$ & High \\ \addlinespace
KL Divergence~\cite{kullback1951information} & $\displaystyle d_{\text{KL}}(p,q) = \sum_{i} p_i \log\frac{p_i}{q_i}$ & $[0, +\infty)$ & $S_{\text{KL}} = \exp(-d_{\text{KL}})$ & High \\ \addlinespace
Euclidean Distance & $\displaystyle d_{\text{Euc}}(p,q) = \sqrt{\sum_{i}(p_i-q_i)^2}$ & $[0, \sqrt{2}]$ & $S_{\text{Euc}} = 1 - \frac{d_{\text{Euc}}}{\sqrt{2}}$ & Low \\ \bottomrule
\end{tabular}
\begin{tablenotes}
\footnotesize
    \item[1] Complexity levels: Low (basic arithmetic operations), Medium (element-wise square roots), High (element-wise logarithmic or exponential operations).
\end{tablenotes}

    \end{threeparttable}
}\vspace{-7mm}
\end{table*}
\begin{table*}[t]
\centering
\caption{Average SGCS performance of Learnware Model Repository with various distributional similarity metrics. CR=1/16 and feedback quantization bits $Q$=4. }
\resizebox{0.98\textwidth}{!}{%
\renewcommand\arraystretch{1}
\begin{threeparttable}
\begin{tabular}{c cccccc cccccc}
\toprule
Similarity Metrics & \multicolumn{6}{c}{LOS} & \multicolumn{6}{c}{NLOS} \\
 \cmidrule(rl){2-7} \cmidrule(rl){8-13} 
 Sampling Range ($\mathrm{m}^2$)&$5\times 5$ &$5\times 10$ &$5\times 20$ &$5\times 30$ &$5\times 40$ &$5\times 50$ &$5\times 5$ &$5\times 10$ &$5\times 20$ &$5\times 30$ &$5\times 40$ &$5\times 50$\\
 \midrule

Cosine ($s_{\cos}$) 
& \textbf{0.886} & 0.855 & 0.839 &0.721  &0.659  &0.599  
&0.324  & \textbf{0.318}  & \textbf{0.319} & \textbf{0.303} & 0.285 &\textbf{0.233}  \\
Bhattacharyya~\cite{bhattacharyya1943measure} ($s_{\text{Bhatt}}$)  
& 0.883 & \textbf{0.857} & \textbf{0.855} &0.746  &0.666  &0.612  
&\textbf{0.327}  & 0.309  & 0.312 & 0.287 & 0.266 &0.209  \\
Hellinger~\cite{hellinger1909neue} ($S_{\text{Hell}}$)  
& 0.883 & \textbf{0.857} & \textbf{0.855} &0.746  &0.666  &0.612  
& \textbf{0.327} & 0.309  & 0.312 & 0.287 & 0.266 & 0.209\\
Total Variation ($S_{\text{TV}}$)  
& 0.884 & 0.856 & 0.854 & 0.721 & 0.669 & 0.625
& 0.324 & 0.307 & 0.302 & 0.300 & 0.283 & 0.226\\
KL~\cite{kullback1951information} ($S_{\text{KL}}$) 
& 0.881 & 0.847 & \textbf{0.855} &\textbf{0.786} &\textbf{0.719} &\textbf{0.657}
& \textbf{0.327} & 0.296 & 0.318 & 0.286 & 0.270 & 0.209\\
Jensen-Shannon~\cite{lin1991divergence} ($S_{\text{JS}}$) 
& 0.883 & \textbf{0.857} & 0.854 &0.746 &0.666 &0.612
&\textbf{0.327} & 0.309 & 0.312 & 0.292 & \textbf{0.289} & 0.209\\
Euclidean ($S_{\text{Euc}}$)   
& 0.873 & 0.856 & 0.759 &0.678 &0.634 &0.528
& 0.325 & 0.316 & 0.315 & 0.299 & 0.278 & 0.217\\\midrule
Original Model Repository
& 0.891 & 0.864 & 0.855 &0.786 &0.737 &0.667
& 0.346 & 0.333 & 0.330 & 0.317 & 0.309 & 0.261\\
Upper Bound 
& 0.915 & 0.881 & 0.903 &0.868 &0.844 &0.848 
& 0.372 & 0.356 & 0.347 & 0.324 & 0.334 & 0.277\\
 
\bottomrule
\end{tabular}
    \end{threeparttable}
}
\label{tab:metric}\vspace{-4mm}
\end{table*}

\subsubsection{Impact of Sample Size on Model Selection Performance}
We investigate how the number of CSI samples affects $\mathbf{pdf}$ estimation accuracy and subsequent model selection performance. In practical deployment, BSs may have limited samples for statistical specification generation, making this analysis essential for understanding real-world applicability.

Fig.~\ref{fig:samplesize_impact} illustrates the impact of sample size on both PDF similarity and SGCS performance. The blue curves with error bars represent PDF similarity, the green curves show average SGCS, and the red dashed lines indicate the benchmark SGCS performance. The results demonstrate that the Learnware Model Repository model exhibits strong robustness to sample size variations, with performance degradation negligible across all tested scenarios. 
For LOS scenes, characterized by concentrated angular energy, even 25 samples yield near-perfect PDF similarity ($>0.99$), and the SGCS performance stabilizes at the benchmark level, confirming that sample size has minimal impact on model selection in such environments. For NLOS scenes, while PDF estimation quality improves progressively with sample size due to diffuse multipath scattering, the impact on SGCS performance remains remarkably small. Remarkably, with only 50 samples, the SGCS already exceeds 98\% of the benchmark, and with 500 samples, the performance gap becomes virtually indistinguishable. These findings confirm that BSs can collect as few as 500 samples to obtain reliable model selection performance regardless of propagation conditions, eliminating the need for extensive data collection in practical deployments.

\subsubsection{Comparison of Distributional Similarity Metrics}
Given that codebook fingerprints are represented as probability distributions, selecting an appropriate similarity metric is critical for accurate learnware retrieval. We systematically evaluate seven metrics: Cosine similarity, Bhattacharyya coefficient~\cite{bhattacharyya1943measure}, Hellinger distance~\cite{hellinger1909neue}, Total Variation, KL divergence~\cite{kullback1951information}, Jensen-Shannon divergence~\cite{lin1991divergence}, and Euclidean distance (details in Tab.~\ref{tab:metrics-formular}). The retrieval performance was measured by average SGCS performance of selected models across various scenarios (6 LOS and 6 NLOS sampling ranges, each with 10 cases), with reference to the Original Model Repository (Top-1 performance, upper bound for Model Repository). 

Tab.~\ref{tab:metric} reveals distinct performance patterns. KL divergence achieves optimal accuracy in LOS large-range environments, matching the repository upper bound. However, it suffers catastrophic failure in NLOS diffuse conditions (0.209 at 5$\times$50 versus 0.261 upper bound). In contrast, Cosine similarity maintains stable performance across all NLOS scenarios (0.324--0.233) without extreme degradation. 
This contrast stems from fundamental differences in handling low-magnitude diffuse distributions. KL divergence's logarithmic form $\sum p_i \log(p_i/q_i)$ amplifies relative errors when absolute probabilities are small (characteristic of NLOS multipath scattering), causing similar diffuse patterns to be misjudged as dissimilar. 
Cosine similarity's $L_2$-normalization $\cos\theta = \frac{\mathbf{p}^T\mathbf{q}}{\|\mathbf{p}\|\|\mathbf{q}\|}$ eliminates sensitivity to absolute energy concentration, focusing on relative angular patterns rather than absolute power levels.

We select Cosine similarity for learnware retrieval due to its robust performance across both LOS concentrated and NLOS scattered distributions, combined with computational efficiency (Low complexity versus High for KL), enabling real-time retrieval in large-scale repositories.

\subsubsection{\textcolor{black}{Multi-carrier OFDM Scenario}}
 
\textcolor{black}{To verify the extensibility of the proposed framework beyond single-carrier systems, we conduct additional experiments on a multi-carrier OFDM scenario with explicit CSI feedback, where the 512-subcarrier channel is transformed via 2D-DFT to the angle-delay domain and pruned to the first 32 delay taps; the real-valued and normalized version of $\mathbf{H} \in \mathbb{C}^{32 \times 32}$ is then fed into the autoencoder. Following the same design principle as the angular-domain codebook fingerprint, we construct a dual-domain statistical specification $\mathbf{pdf}_{\mathrm{joint}} = [\mathbf{pdf}_{\mathrm{angle}};\,\mathbf{pdf}_{\mathrm{delay}}] \in \mathbb{R}^{64}$. Specifically, for each channel sample, we extract the angular domain by projecting each subcarrier's channel vector onto the DFT codebook with $B=5$ and taking a majority vote across all subcarriers to obtain the dominant angular index $\mathrm{idx}_{\mathrm{angle}}$, which filters out subcarrier-level fluctuations and identifies the path consistently observed across the entire bandwidth. 
For the delay domain, since $\mathbf{H}$ is already in the delay domain, we directly average the energy across antennas to obtain the power delay profile, from which the dominant delay index $\mathrm{idx}_{\mathrm{delay}}$ is identified. Aggregating these indices across the dataset yields $\mathbf{pdf}_{\mathrm{angle}} = \mathrm{hist}_{\mathcal{D}}(\mathrm{idx}_{\mathrm{angle}})$ and $\mathbf{pdf}_{\mathrm{delay}} = \mathrm{hist}_{\mathcal{D}}(\mathrm{idx}_{\mathrm{delay}})$.}

\textcolor{black}{
We construct a repository of 200 Learnware models (100 LOS + 100 NLOS) using TF-1 architecture ($L_{\mathrm{TF}}=2$, $feature\_dim$=64, $Embedding\_dim=64$), each trained on 10,000 samples. A TF-4 ($L_{\mathrm{TF}}=12$, $Embedding\_dim=192$) General Model is trained on a mixture of 2 million samples. 
Fig.~\ref{fig:multicarrier_perf} presents the performance on representative LOS/NLOS test cases ($5\mathrm{m}\times 5\mathrm{m}$). The selected Learnware consistently outperforms the General Model and approaches the Original Model Repository, confirming the framework's effectiveness in multi-carrier OFDM systems.
}

\vspace{-5mm}
\section{Conclusion}
\label{conclusion}

This paper proposes a Learnware-based framework for intelligent deployment of CSI feedback models in wireless communication systems. Hosted in a centralized AI data center, the framework supports model reuse across diverse deployment scenarios, reducing training overhead and improving adaptability. To address key limitations of conventional model repositories, such as data privacy concerns, communication costs, and inefficient retrieval, the proposed approach introduces semantic and statistical specifications to enable accurate and lightweight model selection.
The Learnware Model Repository eliminates the need for transmitting raw CSI data and exhaustive model evaluations by leveraging codebook fingerprint distributions for retrieval. An adjustable-speed retrieval strategy ensures scalability, while fine-tuning based on a user-defined threshold enables efficient model adaptation.
Extensive simulations verify that the Learnware model repository consistently outperforms baseline methods.
Compared to the general model with fine-tuning, the Learnware approach achieves higher SGCS performance, \textcolor{black}{with representative gains of 18.8\% in LOS and 57.7\% in NLOS scenarios}, while reducing fine-tuning overhead by 300--1000 samples \textcolor{black}{(depending on propagation conditions)} and 100 training epochs. Furthermore, hierarchical LSH model retrieval and complexity analysis confirms that the retrieval latency is negligible, validating the repository’s practicality for real-time deployment.

\vspace{-4mm}
\bibliographystyle{IEEEtran}
\bibliography{reference}

@techreport{ITU-WP5D1,
	author      = {{ITU-R WP5D}},
	title       = {Framework and overall objectives of the future development of {IMT} for 2030 and beyond},
    institution = {Radio Communication Division of the International Telecommunication Union (ITU-R)},

	year        = {2023},
	month = {Jun.},
	url = {https://www.itu.int/md/R19-WP5D-230612-TD-0905/en}
}

@article{alhammadi2024artificial,
  title={Artificial intelligence in {6G} wireless networks: Opportunities, applications, and challenges},
  author={Alhammadi, Abdulraqeb and Shayea, Ibraheem and El-Saleh, Ayman A and Azmi, Marwan Hadri and Ismail, Zool Hilmi and Kouhalvandi, Lida and Saad, Sawan Ali},
  journal={Int. J. Intell. Syst.},
  volume={2024},
  number={1},
  pages={8845070},
  year={2024},
  month={Mar.},
  publisher={Wiley Online Library}
}

@article{dang2020should,
  title={What should {6G} be?},
  author={Dang, Shuping and Amin, Osama and Shihada, Basem and Alouini, Mohamed-Slim},
  journal={Nature Electron.},
  volume={3},
  number={1},
  pages={20--29},
  year={2020},
  month={Jan.},
  publisher={Nature Publishing Group UK London}
}

@article{ye2024artificial,
  title={Artificial intelligence for wireless physical-layer technologies ({AI4PHY}): A comprehensive survey},
  author={Ye, Neng and Miao, Sirui and Pan, Jianxiong and Ouyang, Qiaolin and Li, Xiangming and Hou, Xiaolin},
  journal={IEEE Trans. Cognit. Commun. Netw.},
  year={2024},
  volume={10},
  number={3},
  pages={729-755},
  month={Jun.},
  publisher={IEEE}
}

@ARTICLE{Hoydis2021Toward,
  author={Hoydis, Jakob and Aoudia, Fayçal Ait and Valcarce, Alvaro and Viswanathan, Harish},
  journal={IEEE Commun. Mag.}, 
  title={Toward a {6G} {AI}-Native Air Interface}, 
  year={2021},
  month={May},
  volume={59},
  number={5},
  pages={76-81}}

@article{marzetta2015massive,
  title={Massive {MIMO}: an introduction},
  author={Marzetta, Thomas L},
  journal={Bell Labs Tech. J.},
  volume={20},
  pages={11--22},
  year={2015},
  month={Mar.},
  publisher={Nokia Bell Labs}
}

@article{eltayeb2014compressive,
  title={Compressive sensing for feedback reduction in {MIMO} broadcast channels},
  author={Eltayeb, Mohammed E and Al-Naffouri, Tareq Y and Bahrami, Hamid Reza},
  journal={IEEE Trans. Commun.},
  volume={62},
  number={9},
  pages={3209--3222},
  year={2014},
  month = {Sept.},
  publisher={IEEE}
}

@INPROCEEDINGS{Ra2015codebook,
  author={Rahman, Md. Saifur and Nam, Young-Han and Zhang, Jianzhong and Seol, Ji-Yun},
  booktitle={Proc. 2015 IEEE Globecom Workshops (GC Wkshps)}, 
  title={Linear Combination Codebook Based {CSI} Feedback Scheme for {FD-MIMO} Systems}, 
  year={2015},
  month={Feb.},
  volume={},
  number={},
  pages={1-6},
  doi={10.1109/GLOCOMW.2015.7414169}}

@article{guo2022overview,
  title={Overview of deep learning-based {CSI} feedback in massive {MIMO} systems},
  author={Guo, Jiajia and Wen, Chao-Kai and Jin, Shi and Li, Geoffrey Ye},
  journal={IEEE Trans. Commun.},
  volume={70},
  number={12},
  pages={8017--8045},
  year={2022},
    month={Dec.},
  publisher={IEEE}
}

@article{wen2018deep,
  title={Deep learning for massive {MIMO CSI} feedback},
  author={Wen, Chao-Kai and Shih, Wan-Ting and Jin, Shi},
  journal={IEEE Wirel. Commun. Lett.},
  volume={7},
  number={5},
  pages={748--751},
  year={2018},
  month = {Oct.},
  publisher={IEEE}
}

@article{an2025deep,
  title={A deep learning-based approach to lightweight {CSI} feedback},
  author={An, Yongli and Lu, Shuoyang and Cai, Haoran and Ji, Zhanlin},
  journal={Phys. Commun.},
  volume={68},
  pages={102538},
  year={2025},
  month = {Feb.},
  publisher={Elsevier}
}

@inproceedings{jiang2024digital,
  title={Digital twin aided massive {MIMO}: {CSI} compression and feedback},
  author={Jiang, Shuaifeng and Alkhateeb, Ahmed},
  booktitle={Proc. ICC 2024-IEEE Int. Conf. Commun.},
  pages={3586--3591},
  year={2024},
  month={Aug.},
  organization={IEEE}
}

@ARTICLE{Chen2025CSI,
  author={Chen, Jie and Hillery, William J. and Mestav, Kursat Rasim and Nuzman, Carl and Saniee, Iraj and Xing, Yunchou},
  journal={IEEE Wirel. Commun.}, 
  title={{CSI} Compression for Massive {MIMO}: Model-Based or Data-Driven?}, 
  year={2025},
  month={Feb.},
  volume={32},
  number={1},
  pages={22-27}
}

@article{liang2024toward,
  title={Toward Better Low-Rate Deep Learning-Based {CSI} Feedback: A Test Channel-Based Approach},
  author={Liang, Xin and Jia, Zhuqing and Gu, Xinyu and Zhang, Lin},
  journal={IEEE Trans. Wirel. Commun.},
  volume={23},
  number={8},
  pages={8773--8786},
  year={2024},
  month = {Aug.},
  publisher={IEEE}
}

@ARTICLE{Shin2024Vector,
  author={Shin, Junyong and Kang, Yujin and Jeon, Yo-Seb},
  journal={IEEE Wirel. Commun. Lett.}, 
  title={Vector Quantization for Deep-Learning-Based {CSI} Feedback in Massive {MIMO} Systems}, 
  year={2024},
  month = {Sep.},
  volume={13},
  number={9},
  pages={2382-2386}}

@ARTICLE{Lin2024Scalable,
  author={Lin, Yu-Chien and Lee, Ta-Sung and Ding, Zhi},
  journal={IEEE Trans. Wirel. Commun.}, 
  title={A Scalable Deep Learning Framework for Dynamic {CSI} Feedback With Variable Antenna Port Numbers}, 
  year={2024},
  month = {Apr.},
  volume={23},
  number={4},
  pages={3102-3116}}

@ARTICLE{Peng2024Learning-Based,
  author={Peng, Zhangjie and Liu, Ruijing and Li, Zhaotian and Pan, Cunhua and Wang, Jiangzhou},
  journal={IEEE Wirel. Commun. Lett.}, 
  title={Deep Learning-Based {CSI} Feedback for {XL-MIMO} Systems in the Near-Field Domain}, 
  year={2024},
  month = {Dec.},
  volume={13},
  number={12},
  pages={3613-3617}}

@ARTICLE{Sun2023Int,
  author={Sun, Jinlong and Zhang, Yibin and Gui, Guan and Zhao, Haitao and Gacanin, Haris and Sari, Hikmet},
  journal={IEEE Trans. Veh. Technol.}, 
  title={Interacting Federated and Transfer Learning-Aided {CSI} Prediction for Intelligent Cellular Networks}, 
  year={2023},
  month = {Dec.},
  volume={72},
  number={12},
  pages={15776-15787}}

@INPROCEEDINGS{Jing2025Source,
  author={Liu, Jingyu and Bouazizi, Mondher and Ohtsuki, Tomoaki},
  booktitle={Proc. 2025 Int. Conf. Comput., Netw. Commun. (ICNC)}, 
  title={Source Channel Model Selection for Downlink {CSI} Feedback Estimation using Transfer Learning in Massive {MIMO} System}, 
  year={2025},
  month={May},
  volume={},
  number={},
  pages={276-281}}

@INPROCEEDINGS{Inoue2023Evalua,
  author={Inoue, Mayuko and Ohtsuki, Tomoaki and Yamamoto, Kohei and Gui, Guan},
  booktitle={Proc. 2023 IEEE 20th Consumer Commun. Netw. Conf. (CCNC)}, 
  title={Evaluation of Source Data Selection for {DTL} Based {CSI} Feedback Method in {FDD} Massive {MIMO} Systems}, 
  year={2023},
  month={Mar.},
  volume={},
  number={},
  pages={182-187}}

@ARTICLE{mingkai2024task,
  author={Chen, Mingkai and Liu, Minghao and Zhang, Zhe and Xu, Zhiping and Wang, Lei},
  journal={China Commun.}, 
  title={Task-oriented semantic communication with foundation models}, 
  year={2024},
  month={Jul.},
  volume={21},
  number={7},
  pages={65-77}}

@article{guo2025prompt,
  title={Prompt-Enabled Large {AI} Models for {CSI} Feedback},
  author={Guo, Jiajia and Cui, Yiming and Wen, Chao-Kai and Jin, Shi},
  journal={IEEE J. Sel. Areas Commun.},
  year={2026},
  mounth={Dec.},
  volume={44},
  number={},
  pages={2654-2668},
}

@ARTICLE{Zhang2024Zone-Specific,
  author={Zhang, Yu and Alkhateeb, Ahmed},
  journal={IEEE Wirel. Commun. Lett.}, 
  title={Zone-Specific {CSI} Feedback for Massive {MIMO}: A Situation-Aware Deep Learning Approach}, 
  year={2024},
  month={Dec.},
  volume={13},
  number={12},
  pages={3320-3324}}

@article{alikhani2024large,
  title={Large wireless model ({LWM}): A foundation model for wireless channels},
  author={Alikhani, Sadjad and Charan, Gouranga and Alkhateeb, Ahmed},
  journal={arXiv preprint arXiv:2411.08872},
  year={2024}
}

@ARTICLE{chen2022implicit,
  author={Chen, Muhan and Guo, Jiajia and Wen, Chao-Kai and Jin, Shi and Li, Geoffrey Ye and Yang, Ang},
  journal={IEEE Trans. Commun.}, 
  title={Deep Learning-Based Implicit {CSI} Feedback in Massive {MIMO}}, 
  year={2022},
  month = {Feb.},
  volume={70},
  number={2},
  pages={935-950}}

@article{guo2025large,
  title={Large {AI} models for wireless physical layer},
  author={Guo, Jiajia and Cui, Yiming and Jin, Shi and Zhang, Jun},
  journal={IEEE Commun. Mag.},
  year={2026},
  month = {May},
  volume={64},
  number={5},
  pages={148-155}
}

@article{zhou2016learnware,
  title={Learnware: on the future of machine learning.},
  author={Zhou, Zhi-Hua},
  journal={Frontiers Comput. Sci.},
  volume={10},
  number={4},
  pages={589--590},
  year={2016},
  month={Aug.}
}

@inproceedings{tan2024beimingwu,
  title={Beimingwu: A learnware dock system},
  author={Tan, Zhi-Hao and Liu, Jian-Dong and Bi, Xiao-Dong and Tan, Peng and Zheng, Qin-Cheng and Liu, Hai-Tian and Xie, Yi and Zou, Xiao-Chuan and Yu, Yang and Zhou, Zhi-Hua},
  booktitle={Proc. 30th ACM SIGKDD Conf. Knowledge Discovery Data Mining},
  pages={5773--5782},
  year={2024},
  month={Aug.}
}

@article{zhou2024learnware,
  title={Learnware: Small models do big},
  author={Zhou, Zhi-Hua and Tan, Zhi-Hao},
  journal={Sci. China Inf. Sci.},
  volume={67},
  number={1},
  pages={112102},
  year={2024},
  month={Oct.},
  publisher={Springer}
}

@inproceedings{zhang2021towards,
  title={Towards enabling learnware to handle unseen jobs},
  author={Zhang, Yu-Jie and Yan, Yu-Hu and Zhao, Peng and Zhou, Zhi-Hua},
  booktitle={Proc. AAAI Conf. Artificial Intelligence},
  volume={35},
  number={12},
  pages={10964--10972},
  year={2021},
  month={May}
}

@inproceedings{lei2024ability,
 author = {Lei, Hao-Yi and Tan, Zhi-Hao and Zhou, Zhi-Hua},
 booktitle = {Proc. 38th Advances Neural Inf. Process. Syst. (NeurIPS)},
 pages = {36471--36513},
 publisher = {Curran Associates, Inc.},
 title = {On the Ability of Developers'  Training Data Preservation of Learnware},
 volume = {37},
 year = {2024},
  month={Dec.}
}

@techreport{3gpp-rp-213599,
  author      = {{3GPP}},
  title       = {New {SI}: Study on Artificial Intelligence ({AI})/Machine Learning ({ML}) for {NR} Air Interface},
  institution = {3GPP},
  type        = {Technical Report},
  number      = {RP-213599},
  year        = {2021},
  month       = {December},
  note        = {3GPP TSG RAN Meeting \#94e, Electronic Meeting}
}

@INPROCEEDINGS{Ahmed2018Explicit,
  author={Ahmed, Rana and Visotsky, Eugene and Wild, Thorsten},
  booktitle={Proc. 22nd Int. ITG Workshop Smart Antennas (WSA)}, 
  title={Explicit {CSI} Feedback Design for {5G} New Radio phase {II}}, 
  year={2018},
  month = {Jun.},
  volume={},
  number={},
  pages={1-5}}

@ARTICLE{baur2024evaluation,
  author={Baur, Michael and Turan, Nurettin and Wallner, Simon and Utschick, Wolfgang},
  journal={IEEE Trans. Machine Learning Commun. Netw.}, 
  title={Evaluation Metrics and Methods for Generative Models in the Wireless {PHY} Layer}, 
  year={2025},
  month={May},
  volume={3},
  number={},
  pages={677-689}}

@ARTICLE{Suh2017Construction,
  author={Suh, Junyeub and Kim, Changhyeon and Sung, Wonjin and So, Jaewoo and Heo, Seo Weon},
  journal={IEEE Commun. Lett.}, 
  title={Construction of a Generalized {DFT} Codebook Using Channel-Adaptive Parameters}, 
  year={2017},
  month = {Jan.},
  volume={21},
  number={1},
  pages={196-199}
  }

@ARTICLE{Li2024Channel,
  author={Li, Yupeng and Li, Gang and Wen, Zirui and Han, Shuangfeng and Gao, Shijian and Liu, Guangyi and Wang, Jiangzhou},
  journal={IEEE Commun. Stand. Mag.}, 
  title={Channel Modeling Aided Dataset Generation For {AI}-Enabled {CSI} Feedback: Advances, Challenges, and Solutions}, 
  year={2024},
  month={Dec.},
  volume={8},
  number={4},
  pages={72-78}}

@Article{QuaDRiGa,
  author  = {S. {Jaeckel} and L. {Raschkowski} and K. {Börner} and L. {Thiele}},
  title   = {{Q}ua{DR}i{G}a: A 3-{D} Multi-Cell Channel Model With Time Evolution for Enabling Virtual Field Trials},
  journal = {IEEE Trans. Antennas Propag.},
  volume  = {62},
  number  = {6},
  pages   = {3242-3256},
  doi     = {10.1109/TAP.2014.2310220},
  year    = {2014},
  month={Jun.},
}

@techreport{3GPP38843,
	author      = {{3GPP TR 38.843}},
	title       = {Study on Artificial Intelligence ({AI})/Machine Learning ({ML}) for {NR} air interface},
    institution = {3rd Generation Partnership Project (3GPP)},
	year        = {2023},
	month = {Dec.},
	url = {https://ftp.3gpp.org//Specs/archive/38_series/38.843/38843-200.zip}
}

@inproceedings{dwivedi2019representation,
  title={Representation similarity analysis for efficient task taxonomy \& transfer learning},
  author={Dwivedi, Kshitij and Roig, Gemma},
  booktitle={Proc. IEEE/CVF Conf. Comput. Vis. Pattern Recognit.(CVPR)},
  pages={12387--12396},
  year={2019}
}

@ARTICLE{han2024ai,
  author={Xiao, Han and Wang, Zhiqin and Li, Dexin and Tian, Wenqiang and Liu, Xiaofeng and Liu, Wendong and Jin, Shi and Shen, Jia and Zhang, Zhi and Yang, Ning},
  journal={China Commun.}, 
  title={{AI} enlightens wireless communication: A transformer backbone for {CSI} feedback}, 
  year={2024},
  month={Dec.},
  volume={21},
  number={12},
  pages={243-256}}

@article{baxter2000model,
  author  = {Jonathan Baxter},
  title   = {A Model of Inductive Bias Learning},
  journal = {J. Artif. Intell. Res. (JAIR)},
  volume  = {12},
  pages   = {149--198},
  year    = {2000},
  month   = {Mar.}
}

@article{maurer2016benefit,
  author  = {Andreas Maurer and Massimiliano Pontil and Bernardino Romera-Paredes},
  title   = {The Benefit of Multitask Representation Learning},
  journal = {J. Artif. Intell. Res. (JAIR)},
  volume  = {17},
  number  = {81},
  pages   = {1--32},
  year    = {2016}
}

@article{hellinger1909neue,
  author  = {E. Hellinger},
  title   = {Neue {B}egr\"{u}ndung der {T}heorie quadratischer {F}ormen von unendlichvielen {V}er\"{a}nderlichen},
  journal = {Crelle's J.},
  volume  = {136},
  pages   = {210--271},
  year    = {1909},
  mouth   = {Dec.}
}

@article{bhattacharyya1943measure,
  author  = {A. Bhattacharyya},
  title   = {On a Measure of Divergence between Two Statistical Populations Defined by Their Probability Distributions},
  journal = {Bull. Calcutta Math. Soc.},
  volume  = {35},
  pages   = {99--110},
  year    = {1943},
  mouth   = {Jan.}
}

@article{lin1991divergence,
  author  = {J. Lin},
  title   = {Divergence Measures Based on the {S}hannon Entropy},
  journal = {IEEE Trans. Inf. Theory},
  volume  = {37},
  number  = {1},
  pages   = {145--151},
  year    = {1991}
}

@article{kullback1951information,
  author  = {S. Kullback and R. A. Leibler},
  title   = {On Information and Sufficiency},
  journal = {Ann. Math. Statist.},
  volume  = {22},
  number  = {1},
  pages   = {79--86},
  year    = {1951}
}

@techreport{3GPP38214,
    type = {Technical Specification},
    key = {3GPP TS 38.214},
    title = {{Technical Specification Group Radio Access Network; NR; Physical layer procedures for data (Release 17)}},
    number = {3GPP TS 38.214 V17.0.0},
    institution = {3rd Generation Partnership Project (3GPP)},
    year = {2022},
    month = mar,
}

@INPROCEEDINGS{Morais2024DatasetSimilarity,
  author={Morais, Joao and Alikhani, Sadjad and Malhotra, Akshay and Hamidi-Rad, Shahab and Alkhateeb, Ahmed},
  booktitle={2024 58th Asilomar Conf. Signals Syst. Comput.},
  title={A Dataset Similarity Evaluation Framework for Wireless Communications and Sensing},
  year={2024},
  pages={1144-1149}
}

@ARTICLE{Lin2025Mobile,
  author={Lin, Xingqin},
  journal={IEEE Commun. Stand. Mag.}, 
  title={A Tale of Two Mobile Generations: {5G-Advanced} and {6G} in {3GPP} Release 20}, 
  year={2025},
  month={Oct.},
  volume={},
  number={},
  pages={1-9}}

@article{huang2025aiml,
  title={{AI/ML} Life Cycle Management for Interoperable {AI} Native {RAN}},
  author={Huang, Chao-Hung and Wen, Chao-Kai and Li, Geoffrey Ye},
  journal={arXiv preprint arXiv:2507.18538},
  year={2025},
  month={jul.}
}

\end{document}